\documentclass[preprint,aps,preprintnumbers]{revtex4}

\usepackage{graphicx,psfrag}
\usepackage{bezier}
\usepackage{latexsym,amssymb}

\begin{document}

\preprint{KANAZAWA-02-39}
\preprint{gr-qc/0212117v3}

\title{The expectation value of the metric operator %
 with respect to Gaussian weave state in loop quantum gravity}

\author{Tomoya Tsushima}

\email{tomoya@hep.s.kanazawa-u.ac.jp}

\affiliation{Institute for Theoretical Physics, Kanazawa University, Kanazawa 920-1192, JAPAN}

\date{\today}

\begin{abstract}
Loop Quantum Gravity is the major candidate of quantum gravity.
It is interesting to consider its continuum limit,
which corresponds to the classical limit.
We consider the Gaussian weave state,
which describes a semi-classical picture.
We calculate the expectation value of
metric operator with respect to this state.
\end{abstract}

\maketitle

\section{Introduction}

In the twenty years,
the canonical quantum gravity has made great progress
\cite{Rovelli:1998yv}.
We call this new stage of theory the loop quantum gravity (LQG).
The quantum state is characterized by closed paths on
three-dimensional space, and is called spin network state.
Its norm is positive definite.
It has been proved non-perturbatively that
the spin network state is an eigen state with respect to
three-dimensional geometrical observables
such as area and volume,
its eigen value is discrete
\cite{Rovelli:1995ge,Loll:1995wt,Frittelli:1996cj,DePietri:1996pj,Ashtekar:1997fb}.
The black hole entropy is calculated by counting the paths of
quantum states crossing the surface of the (classical) event horizon
and is proportional to the area eigenvalue of the surface
\cite{Rovelli:1996dv}.
The non-commutativity of the spatial observable has been also
argued in this framework
\cite{Ashtekar:1998ak}.
Note that the Hamiltonian constraint operator,
which is strongly related to dynamics of the system,
is not been solved so far.
Although the matrix element of it
has been calculated with respect to the spin network state
\cite{Borissov:1997ji,Gaul:2000ba}.

Recently, the ultra high energy cosmic ray (UHECR) events,
whose energy is larger than $10^{11}$ GeV,
have been detected by AGASA experiment
\cite{Takeda:1998ps,Hayashida:2000zr}.
Assume that the UHECR is proton,
it can not travel more than $10^{2}$ Mpc
because it interacts with the cosmic microwave background radiation.
Since there is no nearby sources in our galaxy,
the charged particle cosmic ray spectrum
must obey a high energy energy cutoff, 
witch is called Greisen-Zatsepin-Kuzmin (GZK) cutoff
\cite{Greisen:1966jv,Zatsepin:1966jv}.
However, it conflicts to the data from AGASA.
Then, the AGASA data is a great challenge.

This problem can be solved by taking account of the possible quantum
gravity effects deforming the 
dispersion relation $E^{2}=\vec{p}\;{}^{2}+m^{2}$ due to
the Lorentz invariance
\cite{Amelino-Camelia:1998gz,Ellis:1999rz}.
It is nothing that violation of the relativity
first induced by
\cite{HSato}
with spontaneous breaking down mechanism.
In fact, the semi-classical approximation of quantum gravity 
generally derives
\begin{equation}
 E^{2}=\vec{p}\;^{2}+m^{2}
 +\sum\xi_{n}|\vec{p}\;|^{n}/M_{\mbox{\scriptsize Pl}}^{n-2}
 \label{dispersion}
\end{equation}
where coefficients $\xi_{n}$'s are matter dependent parameters.
In LQG, this relation is derived by
\cite{Gambini:1998it,Alfaro:1999wd,Alfaro:2001rb,Urrutia:2002tr}.
Roughly speaking, LQG is a quantum mechanical system
in the space of $(SU(2))^{n}$.
The wave functional is constructed by $n$ holonomies,
and is square integrable with direct product
of $n$ $SU(2)$ Haar measures.
A holonomy is a path-orderd exponential of $SU(2)$ connection.
In semi-classical approximation,
the expectation velue of the connection is required
as a classical variable instead of the holonomy.
If the holonomy is expanded in power of the connection,
 or of the length of the holonomy,
then spatially non-local terms are generated.
Lorentz symmetry violates.
The mistery of UHECR has been also approached
by enlarging the standard model of elementary particles
without deforming the dispersion relation
\cite{Jain:2000pu,Davoudiasl:2000hv,Hagiwara:2001rm,Ibarra:2002rq,Anchordoqui:2002it}.

More concretely, let us consider
the (densitized inverse) dreibein  operator\
${}^{n\!}\hat{\Theta}^{i}_{I}$,
which plays a role of gravitational part
of Hamiltonian for matter field.
It is expanded with respect to the length of holonomy as
\begin{eqnarray}
 {}^{n\!}\hat{\Theta}^{i}_{I}&=&\frac{4}{i\hbar\kappa}
 \mbox{tr}(\tau^{i}h_{I}\hat{V}^{n}h_{I}^{-1})\nonumber\\
 &\equiv& \frac{2}{i\hbar\kappa}\dot{s}^{a}_{I}[\hat{A}^{i}_{a},\hat{V}^{n}]
 -\frac{1}{i\hbar\kappa}\dot{s}^{a}_{I}\dot{s}^{b}_{I}
 [\partial_{a}\hat{A}^{i}_{b}
 +\frac{1}{2}\epsilon^{ijk}\hat{A}^{j}_{a}\hat{A}^{k}_{b},\hat{V}^{n}]
 +{\cal O}(|\dot{s}^{a}_{I}|^{3}),
 \label{expansion}
\end{eqnarray}
where $\hat{V}$ is the volume operator.
The expectation value of it is estimated in terms of the orders of\
$
 \langle \cdots\hat{A}^{i}_{a}\cdots\rangle\equiv
 \cdots\times(\ell_{\mbox{\scriptsize Pl}}/{\cal L})^{\Upsilon}
 /{\cal L}\times\cdots
$,
where $\cal L$ is a scale that the space looks like continuum for fields.
Because $\hat{A}$ is not defined and state is not decided,
parameter $\Upsilon$ is totally unknown
and the direction and the length of $\dot{s}^{a}$ are also unknown.
These ambiguous situation comes from the
lack of information about the quantum system.

Another discussion,
the coherent state \cite{Thiemann:2000bw} is adopted as a semiclassical state,
and is constructed the cubic graph states
\cite{Sahlmann:2002qj,Sahlmann:2002qk}.
The state has a dimensionless parameter,
which can be smaller than one,
the semi-classical approximation is characterized by this parameter.
The cubic graph is complicated because its vertex is hexavalent.
(The ``valence'' is defined in the next section.)
For weak gravity and calculation simplicity,
the gauge group is reduced from $SU(2)$ to $U(1)^{3}$.


In this paper, we employ a Gaussian weave state
\cite{Corichi:2000gx}, which is superposition of infinite number of states
so that it can describe semiclassical space.
Then we calculate the dreibein operator before series expansion.
Furthermore, the metric operator,
sum of squared dreibein operator,
is calculated in order to determine a mean value of $\dot{s}^{a}$.
We consider the purely gravitational part,
matter contributions are not treated.
In the next section, we explain a essence of LQG we need, and 
determine a notation.
In section III, we introduce a Gaussian weave state,
and calculate the expectation value of the metric operator.
The techniques for spin network calculation is in appendix.

\section{Loop quantum gravity}
\subsection{Real Ashtekar variable}

In LQG, the Einstein-Hilbert action
\begin{eqnarray}
 S = \frac{1}{16\pi G}\int dt\;d^{3}x\;N\sqrt{q}\;
  (R-(K^{a}_{a})^{2}+K_{ab}K^{ab})
\end{eqnarray}
is employed in order to describe the gravitational field. 
$a,b,\cdots$ are spatial indices
and $q_{ab}$ is a three-dimensional metric
and $q=\det(q_{ab})$. 
$R$ and $K_{ab}$ are a three-dimensional Ricci scalar
and an extrinsic curvature, respectivly. 
$G$ is the Newton constant.
The lapse function $N$ corresponds to time-time component 
of four-dimensional metric. Independent variables of this action are,
naively, $q_{ab}$ and its canonical momentum.
Let us introduce a dreibein $\theta^{i}_{a}$ satisfies
the relation $q_{ab}=\theta^{i}_{a}\theta^{i}_{b}$.
The new canonical variables are a dreibein of density weight one\
$
 E^{a}_{i}=\frac{1}{2}\epsilon^{abc}\epsilon_{ijk}\theta^{j}_{b}\theta^{k}_{c}
$\
and a real Ashtekar variable\
\begin{eqnarray}
 A^{i}_{a}=\Gamma^{i}_{a}+\frac{\gamma}{\sqrt{q}\;}E^{b}_{i}K_{ab},
\end{eqnarray}
where $\Gamma^{i}_{a}$ is a spin-connection
that is a function of $E^{b}_{j}$ satisfies the torsionless condition\
$
 \partial_{a}E^{a}_{i}+\epsilon^{ijk}\Gamma^{j}_{a}E^{a}_{k}=0
$\
. The Immirzi parameter $\gamma$ is a real number,
which cannot be determined in the theoretical point of view. 
The indices $i,j,\cdots$ are degrees of freedom of a local internal $SO(3)$.
This canonical pair forms a Poisson bracket
\begin{eqnarray}
 \{A^{i}_{a}(x),E^{b}_{j}(y)\}_{P}
  = \kappa\delta^{b}_{a}\delta^{i}_{j}\delta^{3}(x,y),
\end{eqnarray}
where $\kappa=8\pi\gamma G$, which reproduce the original one
that made by metric and its conjugate.

The real Ashtekar variable $A^{i}_{a}$ behaves as an $SU(2)$ connection form. 
That is, we can regard $A^{i}_{a}$ and $E^{a}_{i}$ as a vector potential
and an electric field in the $SU(2)$ gauge theory, respectively. 
The constraints, which included by this system,
are Hamiltonian constraint\
\begin{eqnarray}
 {\cal H} = \frac{\gamma}{2\kappa\sqrt{q}\;}
  \epsilon^{ijk}E^{a}_{i}E^{b}_{j}
  \left(F^{k}_{ab}-(\gamma^{2}+1)\epsilon^{klm}K^{l}_{a}K^{m}_{b}\right),
 \label{Hamiltonian}
\end{eqnarray}
Gauss constraint\
$
 {\cal G}^{i} = \frac{1}{\kappa}
  (\partial_{a}E^{a}_{i}+\epsilon^{ijk}A^{j}_{a}E^{a}_{k})
$\
and diffeomorphism constraint\
$
 {\cal D}_{a} = \frac{1}{\kappa}E^{b}_{i}F^{i}_{ab}-A^{i}_{a}{\cal G}^{i}
$,
where\
$
 F^{i}_{ab}=\partial_{a}A^{i}_{b}-\partial_{b}A^{i}_{a}
 +\epsilon^{ijk}A^{j}_{a}A^{k}_{b}
$\
is a curvature 2-form of $A^{i}_{a}$, and\
$
 K^{i}_{a}=\frac{1}{\gamma}(A^{i}_{a}-\Gamma^{i}_{a})$.
$\cal H$, ${\cal G}^{i}$ and ${\cal D}_{a}$ are caused by
time reparametrization, $SU(2)$ gauge transformation
and spatial diffeomorphism invariance, respectively.

\subsection{Regularization}

In (\ref{Hamiltonian}), the weight $\sqrt{q}\;$ locates a denominator, 
because the canonical momentum $E^{a}_{i}$ is density weight one
and the integrand must be density weight one.
As an example, the electromagnetic Hamiltonian is
\begin{eqnarray}
 H_{EM}[N]=\int d^{3}x\;\frac{1}{2}N\frac{q_{ab}}{\sqrt{q}\;}
  (\underline{E}^{a}\underline{E}^{b}+\underline{B}^{a}\underline{B}^{b}).
 \label{elemag}
\end{eqnarray}
Since both electric and magnetic field $\underline{E}^{a},\underline{B}^{a}$\
are density weight one, the integrand also has an inverse weight. 
If we naively quantize this, a second order functional derivative
emerges at the same point and the gravitational sector diverges.
Therefore this operator is ill-defined. However, using volume variables
and point splitting methods, we can solve these two types of singularities
\cite{Thiemann:1998aw,Thiemann:1998rt}.

Consider a box, which satisfies $\int_{\mbox{\scriptsize Box}}d^{3}y=\epsilon$,
centered at $x$, and $f_{\epsilon}(x,y)$ is unity if $y$ in the box
and is zero otherwise.
If we take a limit $\epsilon\rightarrow 0$, then\
$\frac{1}{\epsilon}f_{\epsilon}(x,y)\rightarrow\delta^{3}(x,y)$.
The volume variable in the box is
\begin{eqnarray}
 V_{\epsilon}(x)&=&\int d^{3}y\; f_{\epsilon}(x,y)\sqrt{q(y)}\nonumber\\
 &=&\int d^{3}y\; f_{\epsilon}(x,y)
  \sqrt{\left|\frac{1}{6}\epsilon_{abc}\epsilon^{ijk}
  E^{a}_{i}E^{b}_{j}E^{c}_{k}\right|(y)}
\end{eqnarray}
This becomes\
$
  \frac{1}{\epsilon}V_{\epsilon}(x)\rightarrow\sqrt{g(x)}
$\
as $\epsilon\rightarrow 0$. 
Poisson bracket of $A^{i}_{a}$ and $V^{n}$ becomes
a dreibein of density weight ($n-1$):
\begin{eqnarray}
 \left(\frac{\theta^{i}_{a}}{(\sqrt{q})^{1-n}}\right)(x)
 =\lim_{\epsilon\rightarrow 0}\epsilon^{1-n}\frac{2}{\kappa n}
 \{A^{i}_{a}(x),(V_{\epsilon}(x))^{n}\}_{P}
 \label{dens_drei_cl}
\end{eqnarray}
In particular, the density weight becomes a negative
if $n<1$. Let this relation apply to (\ref{elemag}).
This is in the case of $n=1/2$, and there is a factor\
$
 \epsilon^{1/2}\epsilon^{1/2}=\epsilon
$\
in the numerator. Then we should insert a point splitting\
$
 \int d^{3}y\; \frac{1}{\epsilon}f_{\epsilon}(x,y)
$\
to eliminate the $\epsilon$. Thus,
\begin{eqnarray}
 \int d^{3}x\; \frac{q_{ab}}{\sqrt{q}}X^{a}Y^{b}&=&
 \lim_{\epsilon\rightarrow 0}\frac{1}{\epsilon}
  \int d^{3}x\int d^{3}y\; f_{\epsilon}(x,y)
 \left(\frac{\theta^{i}_{a}X^{a}}{(\sqrt{q})^{1/2}}\right)(x)\cdot
  \left(\frac{\theta^{i}_{b}Y^{b}}{(\sqrt{q})^{1/2}}\right)(y)
 \nonumber\\
 &=&\frac{16}{\kappa^{2}}\lim_{\epsilon\rightarrow 0}
  \int d^{3}x\int d^{3}y\; f_{\epsilon}(x,y)
 \nonumber\\
 &&\times
  \left(\{A^{i}_{a},\sqrt{V_{\epsilon}}\}_{P}X^{a}\right)(x)
 \left(\{A^{i}_{b},\sqrt{V_{\epsilon}}\}_{P}Y^{b}\right)(y)
 \label{reg_em_cl}
\end{eqnarray}
This corresponds to the regularized electromagnetic Hamiltonian.
Similarly, The gravitational Hamiltonian (constraint) can be explained
by the commutator with volume variable and point splitting.

We divide a spatial integration into regions specified by $\epsilon$.
We make the region at most includes one vertex of quantum states,
which defined in next subsection.
By this division, if we quantize a regularized variable,
that becomes well-defined operator independently $\epsilon$.
Thus, the limit will be eliminated.

\subsection{Quantum states}

The $SU(2)$ holonomy is a path ordered integral of the connection form
along the smooth path $e$,
\begin{eqnarray}
 h_{e}^{(p)}(A)
  ={\cal P}\exp\left(-\int_{0}^{1}ds\;\dot{e}^{a}(s)\; A^{i}_{a}(e(s))\;
  \tau^{i}_{(p)}\right)
\end{eqnarray}
on the three dimensional space. We call the path $e$ an edge,
and parametrize the orbit $e^{a}$ of the edge by $s\in[0,1]$.\
$\dot{e}^{a}(s)$ is a tangent vector of the edge.
Each of beginning point $e^{a}(0)$ and finalpoint $e^{a}(1)$\
is called vertex.
$\tau^{i}_{(p)}$ is an anti-hermite generator of
$su(2)$ $(p+1)$-dimensional representation, or, equivalently,
spin-$p/2$ representation. 
This integer $p$ is called a color of the edge.
The holonomy changes by gauge transformation
from $h_{e}$ to $g(e(1))h_{e}g^{-1}(e(0))$ for $g(e(1)),g(e(0))\in$ $SU(2)$.

\begin{figure*}[t]
 \psfrag{P0}{$P_{0}$}
 \psfrag{P1}{$P_{1}$}
 \psfrag{P2}{$P_{2}$}
 \psfrag{P3}{$P_{3}$}
 \psfrag{P4}{$P_{4}$}
 \psfrag{i1}{$i_{1}$}
 \psfrag{i2}{$i_{2}$}
 \includegraphics[height=4cm,keepaspectratio,clip]{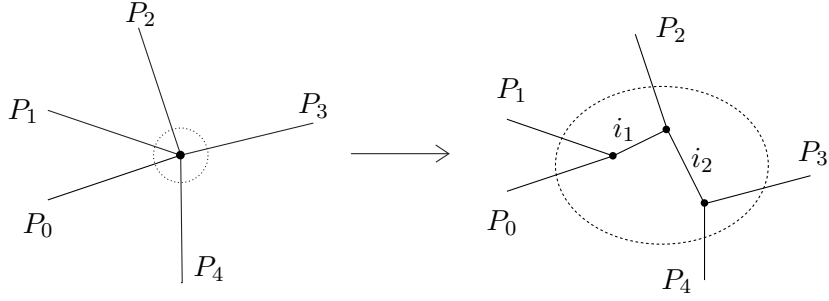} 
\caption{An example of pentavelent vertex.
Five pieces of edges with color $P_{0},\cdots,P_{4}$ have connected
at the vertex. The pentavalent vertex constructed by three virtual trivalent
vertices with two virtual edges of color $i_{2},i_{3}$.}
\label{fig_virtual}
\end{figure*}

Quantum state is characterized by closed graph $\gamma$ constructed by edges.
Edges meet at a vertex. They don't have to connect smoothly.
A vertex which connects $n$ pieces of edges is called $n$-valent vertex.
In trivalent vertex, for example,
if colors of edges are $a,b$ and $c$, respectively
and if $a$ and $b$ are given, $c$ can only take\
$|a-b|,|a-b|+2,\cdots ,a+b-2 $ and $a+b$,
because of $SU(2)$ invariance of quantum states.
$SU(2)$ invariant multi-valent vertex can be constructed by
set of trivalent vertices.
Using $(n-3)$ pieces of `virtual' edge with gauge invariant set
of colors $\vec{i}=\{i_{2},i_{3}\cdots,i_{n-2}\}$,
we can compose the $n$-valent vertex by $(n-2)$ `virtual'
trivalent vertices. The virtual edges have many degrees of freedom
with respect to its colors, we regard as basis of the vertex.
The way of connection between virtual edges
is not unique, but if we fix the basis,
the other way of connection can be described by linear combination
of the basis we chose. This basis, the way of connection between edges,
is called the intertwiner (or intertwining tensor.)
In other words, the intertwiner is a map from tensor product
of incoming edges\
$
 \bigotimes_{e_{I}(1)\cap v}(p(e_{I})+1)
$\
to tensor product of outgoing edges
$\bigotimes_{e_{I}(0)\cap v}(p(e_{I})+1)$.
FIG. \ref{fig_virtual} shows pentavalent vertex, for example.
%
%

The spin network $\{\gamma,\vec{p},\vec{\phi}\}$\
that is a set of a graph $\gamma$,
colors $\vec{p}=\{p(e_{1}),\cdots,p(e_{n})\}$ of edges and intertwiner\
$
 \vec{\phi}=\{\phi_{\vec{i}_{1}}(v_{1}),\cdots,\phi_{\vec{i}_{m}}(v_{m})\}
$\
of vertices $\vec{v}=\{v_{1},\cdots,v_{m}\}$.
It describes a quantum state called a spin network state.
The wave function is explained by
\begin{eqnarray}
 \psi_{\{\gamma,\vec{p},\vec{\phi}\}}(A)
  &=&(h_{e_{1}}\otimes\cdots\otimes h_{e_{n}})
 \cdot(\phi_{\vec{i}_{1}}(v_{1})\otimes\cdots\otimes\phi_{\vec{i}_{m}}(v_{m})).
\end{eqnarray}
Now we define inner product of states of\
$\{\gamma_{1},\vec{p}_{1},\vec{\phi}_{1}\}$\
and $\{\gamma_{2},\vec{p}_{2},\vec{\phi}_{2}\}$.
There is a larger graph\
$
 \gamma=\cup_{I=1}^{n}e_{I}\supset\gamma_{1}\cup\gamma_{2}
$,
then inner product of states can be explained
for $\psi_{\{\gamma_{1},\vec{p},\vec{\phi}\}}$\
and $\psi_{\{\gamma_{2},\vec{p},\vec{\phi}\}}$ on the graph $\gamma$ as
\begin{eqnarray}
 \langle \{\gamma_{1},\vec{p},\vec{\phi}\}|
  \{\gamma_{2},\vec{p},\vec{\phi}\}\rangle &=&
 \int_{(SU(2))^{n}}d\mu(h_{e_{1}})\cdots d\mu(h_{e_{n}})\;
 \overline{f_{\gamma}(h_{e_{1}},\cdots,h_{e_{n}})}\;
  g_{\gamma}(h_{e_{1}},\cdots,h_{e_{n}})\nonumber\\
 &=&\delta_{\gamma_{1},\gamma_{2}}
  \left(\prod_{e\in\gamma}\frac{\delta_{p_{1}(e),p_{2}(e)}}{%
   \Delta_{p_{1}(e)}}\right)\cdot
  \left(\prod_{v\in\gamma}(\phi_{1}(v),\phi_{2}(v))\right)
 \label{norm}
\end{eqnarray}
with the inner product of intertwiners
\begin{eqnarray}
 (\phi_{1}(v),\phi_{2}(v))&=&
  \left(\prod_{\tilde{e}}\frac{
   \delta_{p_{1}(\tilde{e}),p_{2}(\tilde{e})}}{
   \Delta_{p_{1}(\tilde{e})}}\right)
   \left(\prod_{\tilde{v}}\;\theta(p_{1}(e_{1\tilde{v}}),
   p_{1}(e_{2\tilde{v}}),p_{1}(e_{3\tilde{v}}))\right),
 \label{inner}
\end{eqnarray}
where $d\mu$ is $SU(2)$ Haar measure, which normalized $\int_{SU(2)}d\mu=1$.
$\tilde{e},\tilde{v}$ mean a virtual edge and a virtual trivalent vertex
at vertex $v$, respectively.
$e_{1\tilde{v}},e_{2\tilde{v}},e_{3\tilde{v}}$ are edges
(or virtual ones) that connected with the virtual vertex $\tilde{v}$.
The symmetrizer $\Delta_{a}$ and the $\theta$-net $\theta(a,b,c)$ are
defined by (\ref{thesymmetrizer},\ref{thethetanet}).

For simplicity, we denote a spin network state as $\psi_{\gamma}$.
The holonomy $h_{s}$ along to the segment $s$,
is just a product operator that operates to $\psi_{\gamma}$\
adding the edge to the graph, i.e., $h_{s}\psi_{\gamma}=\psi_{s\cup\gamma}$.
The canonical conjugate operator of it is
a left (right) invariant vector\
$(h_{e}\tau^{i})^{A}{}_{B}\partial/\partial(h_{e})^{A}{}_{B}$\
($(\tau^{i}h_{e})^{A}{}_{B}\partial/\partial(h_{e})^{A}{}_{B}$.)
It connects $\tau^{i}$ to the vertex of the edge in the graph.
These operators can be interpreted as
acting on vertices.
Therefore, we set aim to the vertex, and investigate the intertwiner on it.
The normalized intertwiner $\tilde{\phi}_{\vec{i}}$\
of the $n$-valent vertex can be written graphically as
\begin{eqnarray}
 &&\tilde{\phi}_{i_{2},\cdots,i_{n-2}}(P_{0},\cdots,P_{n-1})
 =N_{\vec{i}}(P_{0},\cdots,P_{n-1})
 \begin{array}{c}
 \begin{picture}(130,55)
  \put(15,10){\line(1,0){35}}
  \put(70,10){\line(1,0){25}}
  \put( 5,20){\line(0,1){10}}
  \put(25,10){\line(0,1){20}}
  \put(45,10){\line(0,1){20}}
  \put(85,10){\line(0,1){20}}
  \put(105,20){\line(0,1){10}}
  \put(15,20){\oval(20,20)[bl]}
  \put(95,20){\oval(20,20)[br]}
  \put(25,10){\circle*{3}}
  \put(45,10){\circle*{3}}
  \put(85,10){\circle*{3}}
  \put(55,20){{\scriptsize $\cdots$}}
  \put( 0,35){{\scriptsize $P_{0}$}}
  \put(20,35){{\scriptsize $P_{1}$}}
  \put(40,35){{\scriptsize $P_{2}$}}
  \put(75,35){{\scriptsize $P_{n-2}$}}
  \put(102,35){{\scriptsize $P_{n-1}$}}
  \put(30, 0){{\scriptsize $i_{2}$}}
  \put(70, 0){{\scriptsize $i_{n-2}$}}
  \put( 0,45){{\scriptsize $(e_{0})$}}
  \put(20,45){{\scriptsize $(e_{1})$}}
  \put(40,45){{\scriptsize $(e_{2})$}}
  \put(72,45){{\scriptsize $(e_{n-2})$}}
  \put(100,45){{\scriptsize $(e_{n-1})$}}
 \end{picture}
 \end{array}
\end{eqnarray}
with the normalization factor
\begin{eqnarray}
 N_{\vec{i}}(P_{0},\cdots,P_{n-1})&=&
  \sqrt{\frac{\prod_{x=2}^{n-2}\Delta_{i_{x}}}{%
  \prod_{x=1}^{n-2}\theta(i_{x},P_{x},i_{x+1})}}.
\end{eqnarray}
where $i_{1}=P_{0},\ i_{n-1}=P_{n-1}$.
$e_{0},\cdots,e_{n-1}$ are edges connecting at the vertex,
and $P_{0},\cdots,P_{n-1}$ are its colors respectively.
The inner product (\ref{inner}) of $\tilde{\phi}_{\vec{i}}$ are
given graphically,
\begin{eqnarray}
 (\tilde{\phi}_{\vec{i}}\; ,\ \tilde{\phi}_{\vec{k}})
 &=&N_{\vec{i}}N_{\vec{k}}
 \begin{array}{c}
 \begin{picture}(115,55)
%
  \put(15,10){\line(1,0){20}}
  \put(15,40){\line(1,0){20}}
  \put(55,10){\line(1,0){25}}
  \put(55,40){\line(1,0){25}}
  \put(25,10){\line(0,1){30}}
  \put(65,10){\line(0,1){30}}
  \put(15,25){\oval(20,30)[l]}
  \put(80,25){\oval(20,30)[r]}
  \put(25,10){\circle*{3}}
  \put(25,40){\circle*{3}}
  \put(65,10){\circle*{3}}
  \put(65,40){\circle*{3}}
  \put(42,25){{\scriptsize $\cdots$}}
  \put( 7,23){{\scriptsize $P_{0}$}}
  \put(27,23){{\scriptsize $P_{1}$}}
  \put(67,23){{\scriptsize $P_{n-2}$}}
  \put(92,23){{\scriptsize $P_{n-1}$}}
  \put(29,43){{\scriptsize $k_{2}$}}
  \put(29, 0){{\scriptsize $i_{2}$}}
  \put(48,44){{\scriptsize $k_{n-2}$}}
  \put(48, 0){{\scriptsize $i_{n-2}$}}
 \end{picture}
 \end{array}
 =\prod_{x=2}^{n-2}\delta_{i_{x}}^{k_{x}}.
\end{eqnarray}
If an operator $\hat{X}$ acts to it,
we obtain a matrix element\
$
 X_{\vec{i}}{}^{\vec{k}}=(\tilde{\phi}_{\vec{k}},\hat{X}\tilde{\phi}_{\vec{i}})
$.

\section{The expectation value of the metric operator}
\subsection{Gaussian weave state}

\begin{figure*}[t]
 \psfrag{G1}{$\gamma_{1}$}
 \psfrag{G2}{$\gamma_{2}$}
 \psfrag{p}{\scriptsize $p$}
 \psfrag{0}{\scriptsize 0}
 \includegraphics[height=3cm,keepaspectratio,clip]{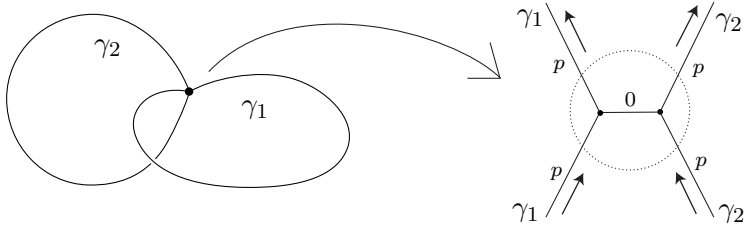} 
\caption{The graph constructed by $\gamma_{1},\;\gamma_{2}$.
The path of the graph is
$
 \gamma^{a}_{1}(0)\rightarrow\gamma^{a}_{1}(1)
 =\gamma^{a}_{2}(0)\rightarrow\gamma^{a}_{2}(1)=\gamma^{a}_{1}(0)
$.
Four tangent vectors\
$\dot{\gamma}^{a}_{1}(0),\dot{\gamma}^{a}_{1}(1),\dot{\gamma}^{a}_{2}(0)$\
and $\dot{\gamma}^{a}_{2}(1)$ are linearly independent.
The vertual vertex can take the color-$0,2,\cdots,2p$,
but we treat the color-0 for simplicity.}
\label{fig_tetra}
\end{figure*}

In LQG, the space is constructed by ``excitation'' of a graph.
Actually, if the graph is not include the multivalent vertex that valence is higher than three, the volume of the space is zero.
Thus, the ``ground state'' is not a flat space.
In order to obtain a flat space,
the graph must include infinite number of vertices.
Therefore, the weave state ${\cal W}=\prod_{v\in R}w_{v}$ in
finite region $R$ in three-dimensional space is defined as following
\cite{Arnsdorf:1999wn,Corichi:2000gx}.
Let us consider the two closed edges $\gamma_{1},\;\gamma_{2}$ crossing
at the vertex $v$.
The wave fuction of color-$p$ is
\begin{eqnarray}
 \Phi_{p}&=&\mbox{tr}(h_{\gamma_{1}}^{(p)}h_{\gamma_{2}}^{(p)})
 =(h^{(p)}_{\gamma_{1}})^{A}{}_{B}\;(h^{(p)}_{\gamma_{2}})^{C}{}_{D}\;
  (\phi_{0}(p,p,p,p))_{A}{}^{B}{}_{C}{}^{D}.
 \label{elementary_state}
\end{eqnarray}
constructed by the edges, as FIG. \ref{fig_tetra}.
%
%
The inner product of $\Phi_{p}$ is normalized.
We take a weave state for each vertex as $\sum_{p=0}^{\infty}c_{p}\Phi_{p}$,
and determine the coefficient as following:
\begin{eqnarray}
 w_{v}=N\exp(-\lambda^{2}(\Phi_{1}-2)^{2}),
 \label{weave_def}
\end{eqnarray}
where $N$, $\lambda$ are normalization factor and any real paremeter,
respectivly.
Using the formula of $SU(2)$ tensor products
\cite{Arnsdorf:1999wn}:
\begin{eqnarray}
 (\Phi_{1})^{n}=\sum_{p}
  \frac{(p+1)\cdot n!}{(\frac{n-p}{2})!(\frac{n+p}{2}+1)!}\Phi_{p}\ ,\ \
  \frac{n-p}{2}=0,1,2,\cdots
\end{eqnarray}
and generating function of Hermite polynomials, exp
$
(-t^{2}+2xt)=\sum H_{n}(x)t^{n}/n!
$,
(\ref{weave_def}) is
\begin{eqnarray}
 w_{v}&=&N\mbox{e}^{-4\lambda^{2}}
  \sum_{n=0}^{\infty}\frac{
   \lambda^{n}H_{n}(2\lambda)}{n!}(\Phi_{1})^{n}\nonumber \\
 &=&N\mbox{e}^{-4\lambda^{2}}
  \sum_{p=0}^{\infty}\sum_{n=0}^{\infty}
  \frac{p+1}{n!(n+p+1)!}\lambda^{2n+p}H_{2n+p}(2\lambda)\cdot \Phi_{p}.
\end{eqnarray}
Then we obtain the coefficient as
\begin{eqnarray}
 c_{p}(\lambda)=N\mbox{e}^{-4\lambda^{2}}
  \sum_{n=0}^{\infty}
  \frac{(p+1)\lambda^{2n+p}H_{2n+p}(2\lambda)}{n!(n+p+1)!}.
 \label{coeff}
\end{eqnarray}
Although the series diverges depending on the value of $\lambda$,
it is at most $\exp(4\lambda^{2})$.
However, in numerical point of view,
it is preferable to converge it.
Moreover, in order to avoid a contribution of higher color,
we employ $\lambda=3/4$ same as
\cite{Corichi:2000gx}.
The norm for each vertex as\
$
 \langle w_{v}|w_{v}\rangle=
 \sum_{p=0}^{\infty}(c_{p}(\frac{3}{4}))^{2}
$.

\subsection{Matrix elements of metric operators}

(\ref{dens_drei_cl}) is rewritten by the holonomies:
\begin{eqnarray}
 {}^{n\!}\Theta^{i}_{I}(v)=-\frac{4}{\kappa n}\mbox{tr}(\tau^{i}h_{I}
  \{h_{I}^{-1},(V_{v})^{n}\}_{P}),
 \label{cl_dreibein}
\end{eqnarray}
where $h_{I}=h_{s_{I}}^{(1)}$,
$s_{I}$ is a segment that endpoint is at $v$.
Let us quantize (\ref{cl_dreibein})
by usual procedure\
$
 \{\cdot,\cdot\}_{P}\rightarrow\frac{1}{i\hbar}[\cdot,\cdot]
$,
then
\begin{eqnarray}
 {}^{n\!}\hat{\Theta}^{i}_{I}(v)&=&-\frac{4}{i\hbar\kappa n}
  \mbox{tr}(\tau^{i}h_{I}[h_{I}^{-1},\hat{V}_{v}^{n}])
 =\frac{4}{i\hbar\kappa n}\mbox{tr}(\tau^{i}h_{I}\hat{V}_{v}^{n}h_{I}^{-1}).
\end{eqnarray}
${}^{n}\!\hat{\Theta}^{i}_{I}(v)$ is equal to\
$\hat{Q}^{i}_{e_{I}}(v,n)$ times $-8i\hbar$\
defined in
\cite{Sahlmann:2002qj}.
We operate it to $w_{v}$.
First, $h_{I}^{-1}$ acts to a trivalent vertex $\tilde{\phi}_{k}(p,p,p,p)$:
\begin{eqnarray}
 h^{-1}_{0}\;\tilde{\phi}_{k}(p,p,p,p)&=&N_{k}(p)\cdot
 \begin{array}{c}
 \begin{picture}(75,45)
%
  \put(20,10){\line(1,0){40}}
  \put(10,20){\line(0,1){10}}
  \put(30,10){\line(0,1){20}}
  \put(50,10){\line(0,1){20}}
  \put(70,20){\line(0,1){10}}
  \put(20,20){\oval(20,20)[bl]}
  \put(60,20){\oval(20,20)[br]}
  \put(30,10){\circle*{3}}
  \put(50,10){\circle*{3}}
  \put(8,35){{\scriptsize $p$}}
  \put(28,35){{\scriptsize $p$}}
  \put(48,35){{\scriptsize $p$}}
  \put(68,35){{\scriptsize $p$}}
  \put(38, 0){{\scriptsize $k$}}
  \put(19,19){\oval(28,28)[bl]}
  \put(19,5){\line(1,0){9}}
  \put(5,19){\line(0,1){11}}
  \put(-1,15){{\scriptsize $1$}}
 \end{picture}
 \end{array}\nonumber\\
 &=&N_{k}(p)
 \sum_{q=\pm 1}\alpha_{q}(p)
 \begin{array}{c}
 \begin{picture}(105,45)
%
  \put(30,10){\line(1,0){60}}
  \put(20,20){\line(0,1){10}}
  \put(60,10){\line(0,1){20}}
  \put(80,10){\line(0,1){20}}
  \put(100,20){\line(0,1){10}}
  \put(30,20){\oval(20,20)[bl]}
  \put(90,20){\oval(20,20)[br]}
  \put(60,10){\circle*{3}}
  \put(80,10){\circle*{3}}
  \put(18,35){\scriptsize $p$}
  \put(58,35){\scriptsize $p$}
  \put(78,35){\scriptsize $p$}
  \put(98,35){\scriptsize $p$}
  \put(68, 2){\scriptsize $k$}
  \put(10,23){\line(1,0){10}}
  \put(40, 0){\line(0,1){10}}
  \put(40,10){\circle*{3}}
  \put(20,23){\circle*{3}}
  \put( 2,22){\scriptsize $1$}
  \put(33, 0){\scriptsize $1$}
  \put(24,15){\scriptsize $p+q$}
  \put(47, 3){\scriptsize $p$}
 \end{picture}
 \end{array},
\end{eqnarray}
where $\alpha_{+1}(p)=1,\ \alpha_{-1}=-\frac{p}{p+1}$ and\
$N_{k}(p)=N_{k}(p,p,p,p)$.
It can be regarded as a pentavalent vertex
for volume operator. Denote
\begin{eqnarray}
 \tilde{\phi}_{ij}(\check{1},q,p,p,p)&=&N_{ij}(\check{1},q,p,p,p)
 \begin{array}{c}
 \begin{picture}(90,50)
  \put(15,10){\line(1,0){60}}
  \put( 5,20){\line(0,1){10}}
  \put(25,10){\line(0,1){20}}
  \put(45,10){\line(0,1){20}}
  \put(65,10){\line(0,1){20}}
  \put(85,20){\line(0,1){10}}
  \put(15,20){\oval(20,20)[bl]}
  \put(75,20){\oval(20,20)[br]}
  \put(25,10){\circle*{3}}
  \put(45,10){\circle*{3}}
  \put(65,10){\circle*{3}}
  \put( 3,35){{\scriptsize $\check{1}$}}
  \put(23,35){{\scriptsize $q$}}
  \put(43,35){{\scriptsize $p$}}
  \put(63,35){{\scriptsize $p$}}
  \put(83,35){{\scriptsize $p$}}
  \put(30, 3){{\scriptsize $i$}}
  \put(53, 3){{\scriptsize $j$}}
  \put(20,43){{\scriptsize $(e_{0})$}}
  \put(40,43){{\scriptsize $(e_{1})$}}
  \put(60,43){{\scriptsize $(e_{2})$}}
  \put(80,43){{\scriptsize $(e_{3})$}}
 \end{picture}
 \end{array},
\end{eqnarray}
where the check $\check{1}$ means that
the volume operator does not operate to it.
By (\ref{volume_element}),
matrix element of volume operator is found,
$
 \hat{V}^{n}\tilde{\phi}_{ij}
 =\sum_{s,t}V^{n}{}_{ij}{}^{st}\tilde{\phi}_{st}
$,
then
\begin{eqnarray}
 \hat{V}^{n}h_{0}^{-1}\; \tilde{\phi}_{k}&=&
  \sum_{q=\pm 1}\sum_{s,t}\alpha_{q}(p)N_{k}(p)
  \left(\frac{N_{st}V^{n}{}_{pk}{}^{st}}{N_{pk}}\right)(\check{1},p+q,p,p,p)
 \nonumber\\
 &&\times
 \begin{array}{c}
 \begin{picture}(70,50)
  \put(30,10){\line(1,0){60}}
  \put(20,20){\line(0,1){10}}
  \put(60,10){\line(0,1){20}}
  \put(80,10){\line(0,1){20}}
  \put(100,20){\line(0,1){10}}
  \put(30,20){\oval(20,20)[bl]}
  \put(90,20){\oval(20,20)[br]}
  \put(60,10){\circle*{3}}
  \put(80,10){\circle*{3}}
  \put(18,35){\scriptsize $p$}
  \put(58,35){\scriptsize $p$}
  \put(78,35){\scriptsize $p$}
  \put(98,35){\scriptsize $p$}
  \put(68, 3){\scriptsize $t$}
  \put(10,23){\line(1,0){10}}
  \put(40, 0){\line(0,1){10}}
  \put(40,10){\circle*{3}}
  \put(20,23){\circle*{3}}
  \put( 2,22){\scriptsize $1$}
  \put(33, 0){\scriptsize $1$}
  \put(26,15){\scriptsize $p+q$}
  \put(47, 3){\scriptsize $s$}
 \end{picture}
 \end{array}
\end{eqnarray}
Therefore, the action of dreibein operator with density weight $(n-1)$\
is obtained as
\begin{eqnarray}
 {}^{n\!}\hat{\Theta}^{i}_{0}\; \tilde{\phi}_{k}
 &=&-\frac{4}{i\hbar\kappa n}\sum_{q=\pm 1}
  \sum_{s,t}
  \alpha_{q}(p)N_{k}(p)
  \left(\frac{N_{st}V^{n}{}_{pk}{}^{st}}{N_{pk}}\right)(\check{1},p+q,p,p,p)
  \nonumber\\
 &&\hspace{5em}\times \frac{\mbox{Tet}\left[%
 \begin{array}{ccc}
  s & p & p+q\\
  1 & 1 & 2
 \end{array}%
 \right]}{\theta(s,p,2)}
 \cdot\!\!
 \begin{array}{l}
 \begin{picture}(100,40)
%
  \put(8,22){\circle{10}}
  \put(6,19){{\scriptsize $i$}}
  \put(12,27){\oval(8,10)[tl]}
  \put(12,17){\oval(8,10)[bl]}
  \put(12,22){\oval(12,20)[r]}
  \put(18,23){\line(1,0){17}}
  \put(18,23){\circle*{3}}
  \put(24,25){{\scriptsize 2}}
  \put(35,23){\circle*{3}}
  \put(45,10){\line(1,0){40}}
  \put(35,20){\line(0,1){10}}
  \put(55,10){\line(0,1){20}}
  \put(75,10){\line(0,1){20}}
  \put(95,20){\line(0,1){10}}
  \put(45,20){\oval(20,20)[bl]}
  \put(85,20){\oval(20,20)[br]}
  \put(55,10){\circle*{3}}
  \put(75,10){\circle*{3}}
  \put(33,35){{\scriptsize $p$}}
  \put(53,35){{\scriptsize $p$}}
  \put(73,35){{\scriptsize $p$}}
  \put(93,35){{\scriptsize $p$}}
  \put(63, 3){{\scriptsize $t$}}
  \put(40,15){{\scriptsize $s$}}
 \end{picture}
 \end{array}.
 \label{drei_matrix}
\end{eqnarray}
In right hand side of (\ref{drei_matrix}),
$su(2)$ generator connects the intertwiner.
Since it is not $SU(2)$ invariant, the expectation value becomes zero.
Thus we operate it two times, that is, metric operator:
\begin{eqnarray}
 {}^{n\!}\hat{q}(s_{I},s_{J})(v)=\frac{1}{2}
  ({}^{n\!}\hat{\Theta}_{I}^{i}(v)\;{}^{n\!}\hat{\Theta}_{J}^{i}(v)
  +{}^{n\!}\hat{\Theta}_{J}^{i}(v)\;{}^{n\!}\hat{\Theta}_{I}^{i}(v)).
 \label{metric_op}
\end{eqnarray}
It operates to two vertices but it is same points,
because its expectation value vanishes by spin network calculation
when it acts different points.
In generally, dreibein operator is non-commutative:\
$[\hat{\Theta}^{i}_{I}(v),\hat{\Theta}^{j}_{J}(v)]\neq 0$,
thus we take (\ref{metric_op}) to be symmetric explicitly.
Let the graphical part of right hand side of (\ref{drei_matrix})
denote as
\begin{eqnarray}
 f^{i}_{st}=
 \begin{array}{l}
 \begin{picture}(110,55)
%
  \put(8,22){\circle{10}}
  \put(6,19){{\scriptsize $i$}}
  \put(12,27){\oval(8,10)[tl]}
  \put(12,17){\oval(8,10)[bl]}
  \put(12,22){\oval(12,20)[r]}
  \put(18,23){\line(1,0){17}}
  \put(18,23){\circle*{3}}
  \put(24,25){{\scriptsize 2}}
  \put(35,23){\circle*{3}}
  \put(45,10){\line(1,0){40}}
  \put(35,20){\line(0,1){10}}
  \put(55,10){\line(0,1){20}}
  \put(75,10){\line(0,1){20}}
  \put(95,20){\line(0,1){10}}
  \put(45,20){\oval(20,20)[bl]}
  \put(85,20){\oval(20,20)[br]}
  \put(55,10){\circle*{3}}
  \put(75,10){\circle*{3}}
  \put(33,35){{\scriptsize $p$}}
  \put(53,35){{\scriptsize $p$}}
  \put(73,35){{\scriptsize $p$}}
  \put(93,35){{\scriptsize $p$}}
  \put(63, 3){{\scriptsize $t$}}
  \put(40,15){{\scriptsize $s$}}
  \put(31,43){{\scriptsize $(e_{0})$}}
  \put(51,43){{\scriptsize $(e_{1})$}}
  \put(71,43){{\scriptsize $(e_{2})$}}
  \put(91,43){{\scriptsize $(e_{3})$}}
 \end{picture}
 \end{array}.
 \label{dreibein_to_state}
\end{eqnarray}
We operate a dreibein operator to it. 
There is two cases, $I=J$ and $I\neq J$, in the operation.
In the case of $I=J=0$, the action of $h_{0}^{-1}$ is
\begin{eqnarray}
 h_{0}^{-1}\; f^{i}_{st}&=&
 \begin{array}{l}
 \begin{picture}(105,45)
%
  \put(8,22){\circle{10}}
  \put(6,19){{\scriptsize $i$}}
  \put(12,27){\oval(8,10)[tl]}
  \put(12,17){\oval(8,10)[bl]}
  \put(12,22){\oval(12,20)[r]}
  \put(18,23){\line(1,0){22}}
  \put(18,23){\circle*{3}}
  \put(24,25){{\scriptsize 2}}
  \put(40,23){\circle*{3}}
  \put(50,10){\line(1,0){40}}
  \put(40,20){\line(0,1){10}}
  \put(60,10){\line(0,1){20}}
  \put(80,10){\line(0,1){20}}
  \put(100,20){\line(0,1){10}}
  \put(50,20){\oval(20,20)[bl]}
  \put(90,20){\oval(20,20)[br]}
  \put(60,10){\circle*{3}}
  \put(80,10){\circle*{3}}
  \put(38,35){{\scriptsize $p$}}
  \put(58,35){{\scriptsize $p$}}
  \put(78,35){{\scriptsize $p$}}
  \put(98,35){{\scriptsize $p$}}
  \put(68, 3){{\scriptsize $t$}}
  \put(45,15){{\scriptsize $s$}}
  \put(50,20){\oval(30,30)[bl]}
  \put(50, 5){\line(1,0){10}}
  \put(35,25){\line(0,1){5}}
  \put(30, 5){{\scriptsize $1$}}
 \end{picture}
 \end{array}
 =\sum_{v=\pm 1}\alpha_{v}(s)
 \begin{array}{l}
 \begin{picture}(120,65)
%
  \put(8,37){\circle{10}}
  \put(6,34){{\scriptsize $i$}}
  \put(12,42){\oval(8,10)[tl]}
  \put(12,32){\oval(8,10)[bl]}
  \put(12,37){\oval(12,20)[r]}
  \put(18,38){\line(1,0){17}}
  \put(18,38){\circle*{3}}
  \put(24,40){{\scriptsize 2}}
  \put(35,38){\circle*{3}}
  \put( 45,10){\line(1,0){60}}
  \put( 35,20){\line(0,1){30}}
  \put( 75,10){\line(0,1){40}}
  \put( 95,10){\line(0,1){40}}
  \put(115,20){\line(0,1){30}}
  \put( 45,20){\oval(20,20)[bl]}
  \put(105,20){\oval(20,20)[br]}
  \put( 75,10){\circle*{3}}
  \put( 95,10){\circle*{3}}
  \put( 33,55){\scriptsize $p$}
  \put( 73,55){\scriptsize $p$}
  \put( 93,55){\scriptsize $p$}
  \put(113,55){\scriptsize $p$}
  \put( 83, 3){\scriptsize $t$}
  \put( 38,15){\tiny $|s+v|$}
  \put( 38,30){\scriptsize $s$}
  \put( 62,12){\scriptsize $s$}
  \put(28,25){\line(1,0){7}}
  \put(55, 3){\line(0,1){7}}
  \put(22,22){\scriptsize $1$}
  \put(49, 0){\scriptsize $1$}
  \put(35,25){\circle*{3}}
  \put(55,10){\circle*{3}}
 \end{picture}
 \end{array}.
\end{eqnarray}
The volume operator regards it as\
$\tilde{\phi}_{st}(\check{1},|s+v|,p,p,p)$.
Therefore, the matrix element of the metric operator of same direction is
\begin{eqnarray}
 {}^{n\!}\hat{q}(s_{0},s_{0})(v)\;
  \tilde{\phi}_{k}&=&
 -\frac{8}{(\hbar\kappa n)^{2}}\sum_{q=\pm 1}
  \sum_{s,t}\sum_{v=\pm 1}\sum_{m}
 \nonumber\\
 &&\times
  \alpha_{q}(p)\alpha_{v}(s)
  \left(\frac{N_{k}}{N_{m}}\right)(p)
 \nonumber\\
 &&\times
  \left(\frac{N_{st}V^{n}{}_{pk}{}^{st}}{%
   N_{pk}}\right)(\check{1},p+q,p,p,p)
  \left(\frac{N_{pm}V^{n}{}_{st}{}^{pm}}{%
   N_{st}}\right)(\check{1},|s+v|,p,p,p)
 \nonumber\\
 &&\times
 \frac{\mbox{Tet}\left[%
 \begin{array}{ccc}
  s & p & p+q\\
  1 & 1 & 2
 \end{array}%
 \right]\mbox{Tet}\left[%
 \begin{array}{ccc}
  p & s & |s+v|\\
  1 & 1 & 2
 \end{array}%
 \right]}{\Delta_{p}\theta(s,p,2)}\; \tilde{\phi}_{m}\nonumber\\
 &=&\sum_{m}{}^{n\!}q(s_{0},s_{0})_{(p)km}\; \tilde{\phi}_{m}.
\end{eqnarray}
Similarly, all
$
 {}^{n\!}\hat{q}(s_{I},s_{J})(v)
$\
can also be obtained.

Concretely, we show the expectation value of ${}^{n\!}\hat{q}(s_{I},s_{J})$\
numerically.
In (\ref{coeff}), let the norm be normalize\
$\sum_{p}(c_{p})^{2}(\lambda)=1$,
and we set a parameter $\lambda=3/4$,
then $(c_{0})^{2}=0.414892,\ (c_{1})^{2}=0.482013$\
and $(c_{2})^{2}=0.0972374$.
Since the coefficients of
higher color more than $p=3$ are negligible
such as $(c_{3})^{2}=1.28919\times 10^{-6}$,
it is sufficient to evaluate the contribution of $p=1,2$.
Then we obtain
\begin{equation}
 \begin{array}{rl}
 {}^{1\!}q(s_{0},s_{0})_{00}&=0.150\hbar\kappa,\\
 {}^{1\!}q(s_{1},s_{1})_{00}&=0.131\hbar\kappa,\\
 {}^{1\!}q(s_{2},s_{2})_{00}&=0.265\hbar\kappa,\\
 {}^{1\!}q(s_{3},s_{3})_{00}&=0.0980\hbar\kappa.
 \end{array}
\end{equation}
The state (\ref{elementary_state}) is not symmetric
with respect to edges, the values of length squared of tangent vectors 
at $v$ are different each other.
We can assign a distance between vertices
to an average of $\dot{s}_{I}$'s length,\
$
 \frac{1}{4}\sum_{I}\sqrt{{}^{1\!}q(s_{I},s_{I})_{00}}
 = 0.394(\hbar\kappa)^{1/2}
$.
The some angles between $\dot{s}^{a}_{I}$'s,
which defined by
$
 \theta(s_{I},s_{J})=\cos^{-1}
 ({}^{1\!}q(s_{I},s_{J})_{00}/
 (\sqrt{{}^{1\!}q(s_{I},s_{I})_{00}}
  \sqrt{{}^{1\!}q(s_{J},s_{J})_{00}}))
$,
are also obtained as
\begin{equation}
 \begin{array}{rl}
 \theta(s_{0},s_{1}) &= 160^{\circ},\\
 \theta(s_{0},s_{2}) &= 74.6^{\circ},\\
 \theta(s_{0},s_{3}) &= 83.0^{\circ}.
 \end{array}
\end{equation}
Thus, the direction of $\dot{s}^{a}_{I}$ is not symmetric in this state.

\section{Conclusion}

We calculated that the expectation value of the metric operator
with respect to Gaussian weave state.
It can be identified to the length squared
of the edge $|\dot{s}^{a}_{I}|^{2}$,
and it determines the scale of this system.
We found that the diagonal element of the metric is surely positive definite.
This result seems non-trivial at first sight,
because the metric operator
\begin{eqnarray}
 {}^{n\!}\hat{q}(s_{I},s_{I})&=&
 \frac{4}{(\hbar\kappa n)^{2}}\left(
 2\mbox{tr}(h_{I}\hat{V}^{2n}h_{I}^{-1})
  - \left(\mbox{tr}(h_{I}\hat{V}^{n}h_{I}^{-1})\right)^{2}
 \right)
\end{eqnarray}
has a negative sign in second term,
and the matrix element of the volume operator is not positive definite.
(The eigenvalue of it is positive definite.)
However, 
by the spin network calculation,
\begin{eqnarray}
 ||{}^{n\!}\hat{\Theta}^{i}_{0}\tilde{\phi}_{k}||^{2}
 &=&
 \sum_{s,t}(\mbox{positive number})\times
 \begin{array}{l}
 \begin{picture}(110,65)
%
  \put( 8,22){\circle{10}}
  \put( 6,19){{\scriptsize $i$}}
  \put(12,27){\oval(8,10)[tl]}
  \put(12,17){\oval(8,10)[bl]}
  \put(12,22){\oval(12,20)[r]}
  \put(18,23){\line(1,0){17}}
  \put(18,23){\circle*{3}}
  \put(24,25){{\scriptsize 2}}
  \put( 8,47){\circle{10}}
  \put( 6,44){{\scriptsize $i$}}
  \put(12,52){\oval(8,10)[tl]}
  \put(12,42){\oval(8,10)[bl]}
  \put(12,47){\oval(12,20)[r]}
  \put(18,48){\line(1,0){17}}
  \put(18,48){\circle*{3}}
  \put(24,50){{\scriptsize 2}}
  \put(35,23){\circle*{3}}
  \put(35,48){\circle*{3}}
  \put(45,10){\line(1,0){40}}
  \put(45,58){\line(1,0){40}}
  \put(35,20){\line(0,1){28}}
  \put(55,10){\line(0,1){48}}
  \put(75,10){\line(0,1){48}}
  \put(95,20){\line(0,1){28}}
  \put(45,20){\oval(20,20)[bl]}
  \put(85,20){\oval(20,20)[br]}
  \put(45,48){\oval(20,20)[tl]}
  \put(85,48){\oval(20,20)[tr]}
  \put(55,10){\circle*{3}}
  \put(75,10){\circle*{3}}
  \put(55,58){\circle*{3}}
  \put(75,58){\circle*{3}}
  \put(38,35){{\scriptsize $p$}}
  \put(58,35){{\scriptsize $p$}}
  \put(78,35){{\scriptsize $p$}}
  \put(98,35){{\scriptsize $p$}}
  \put(63, 3){{\scriptsize $t$}}
  \put(63,50){{\scriptsize $t$}}
  \put(40,15){{\scriptsize $s$}}
  \put(40,51){{\scriptsize $s$}}
 \end{picture}
 \end{array}\nonumber\\
 &=& -(\mbox{positive number})\times\mbox{tr}(\tau^{i}\tau^{i})
 \geq 0,\ \ (\mbox{no-sum with respect to }i).
\end{eqnarray}
That is,\
$
 \langle w_{v}| {}^{n\!}\hat{q}(s_{I},s_{I})w_{v}\rangle
 =\sum_{i}||{}^{n\!}\hat{\Theta}^{i}_{I}w_{v}||^{2} \geq 0
$.



\begin{acknowledgments}
I would like to thank Ken-Ichi Aoki,
for helpful comments and useful advice.
\end{acknowledgments}
\appendix
\section{Spin network calculation}

\subsection{Spin network and recoupling theory}

In this subsection, we refered to 
\cite{DePietri:1996pj}.
The $SU(2)$ invariant tensor, a matrix $X^{A}{}_{B}$\
and its trace are graphically written as
\begin{eqnarray}
  \delta^{A}_{B}=
  \begin{array}{l}
    \begin{picture}(20,20)
      \put(10,0){\line(0,1){20}}
      \put(2,15){{\scriptsize $A$}}
      \put(12,0){{\scriptsize $B$}}
    \end{picture}
  \end{array}&,&\ \
  \delta^{A}_{D}\delta^{B}_{C}=-
  \begin{array}{l}
    \begin{picture}(20,20)
      \put(7,0){\line(1,2){10}}
      \put(17,0){\line(-1,2){10}}
      \put(0,15){{\scriptsize $A$}}
      \put(18,15){{\scriptsize $B$}}
      \put(0,0){{\scriptsize $C$}}
      \put(18,0){{\scriptsize $D$}}
    \end{picture}
  \end{array}\hspace{2em},\hspace{2em}
  i\epsilon^{AB}=
  \begin{array}{l}
    \begin{picture}(20,20)
      \put(12,5){\oval(10,10)[b]}
      \put(7,5){\line(0,1){10}}
      \put(17,5){\line(0,1){10}}
      \put(0,15){{\scriptsize $A$}}
      \put(18,15){{\scriptsize $B$}}
    \end{picture}
  \end{array}\nonumber\\
  i\epsilon_{AB}=
  \begin{array}{l}
    \begin{picture}(20,20)
      \put(12,15){\oval(10,10)[t]}
      \put( 7, 5){\line(0,1){10}}
      \put(17, 5){\line(0,1){10}}
      \put( 0, 0){\scriptsize $A$}
      \put(18, 0){\scriptsize $B$}
    \end{picture}
  \end{array}&,&\ \
 X^{A}{}_{B}=
 \begin{array}{l}
 \begin{picture}(25,30)
%
 \put( 8,10){\framebox(10,10)}
 \put( 9,12){\scriptsize $X$}
 \put(13,20){\line(0,1){10}}
 \put(13, 0){\line(0,1){10}}
 \put( 5,25){\scriptsize $A$}
 \put(16, 1){\scriptsize $B$}
 \end{picture}
 \end{array}\hspace{2em},\hspace{2em}
 \mbox{tr}X=-
 \begin{array}{l}
 \begin{picture}(25,30)
%
 \put( 3,10){\framebox(10,10)}
 \put( 4,12){\scriptsize $X$}
 \put(14,10){\oval(10,10)[b]}
 \put(14,20){\oval(10,10)[t]}
 \put(19,10){\line(0,1){10}}
 \end{picture}
 \end{array}.
\end{eqnarray}
Since we add the minus signature to the crossing line,
a symmetric tensor becomes anti-symmetric line.
We derive the binor identity
\begin{eqnarray}
\begin{array}{l}
\begin{picture}(20,20)
 \put(5,0){\line(0,1){20}}
 \put(15,0){\line(0,1){20}}
\end{picture}
\end{array}
 +
\begin{array}{l}
\begin{picture}(20,20)
 \put(5,0){\line(1,2){10}}
 \put(15,0){\line(-1,2){10}}
\end{picture}
\end{array}
 +
\begin{array}{l}
\begin{picture}(20,20)
 \put(10,20){\oval(10,14)[b]}
 \put(10,0){\oval(10,14)[t]}
 \end{picture}
\end{array} = 0.
\end{eqnarray}
Thus, the lines behave as knots satisfying the Reidemeister moves
O, I, II and III:
\begin{eqnarray}
 \mbox{O}&:&
 \begin{array}{l}
 \begin{picture}(30,20)
%
  \put(10,10){\oval(10,10)[t]}
  \put(20,10){\oval(10,10)[b]}
  \put( 5, 0){\line( 0, 1){10}}
  \put(25,10){\line( 0, 1){10}}
 \end{picture}
 \end{array}
 =
 \begin{array}{l}
 \begin{picture}(10,20)
%
 \put(5,0){\line(0,1){20}}
 \end{picture}
 \end{array},\nonumber\\
 \mbox{I}&:&
 \begin{array}{l}
 \begin{picture}(30,20)
%
  \put(15, 0){\line( 1, 1){10}}
  \put(25, 0){\line(-1, 1){10}}
  \put(20,10){\oval(10,20)[t]}
 \end{picture}
 \end{array}
 =
 \begin{array}{l}
 \begin{picture}(20,20)
%
  \put(10,15){\oval(10,10)[t]}
  \put( 5,0){\line(0,1){15}}
  \put(15,0){\line(0,1){15}}
 \end{picture}
 \end{array},\nonumber\\
 \mbox{II}&:&
 \begin{array}{l}
 \begin{picture}(30,20)
%
 \put(10,12){\oval(20,10)[r]}
 \put(20, 8){\oval(20,10)[l]}
 \put( 0, 7){\line( 1, 0){10}}
 \put( 0,17){\line( 1, 0){10}}
 \put(20, 3){\line( 1, 0){10}}
 \put(20,13){\line( 1, 0){10}}
 \end{picture}
 \end{array}
 =
 \begin{array}{l}
 \begin{picture}(30,20)
%
 \put( 0,12){\oval(20,10)[r]}
 \put(30, 8){\oval(20,10)[l]}
 \end{picture}
 \end{array},\nonumber\\
 \mbox{III}&:&
 \begin{array}{l}
 \begin{picture}(30,20)
%
  \put( 5, 0){\line( 1, 1){20}}
  \put(25, 0){\line(-1, 1){20}}
  \put(18,20){\oval( 6,10)[lb]}
  \put(18,10){\oval( 6,10)[r]}
  \put(18, 0){\oval( 6,10)[lt]}
 \end{picture}
 \end{array}
 =
 \begin{array}{l}
 \begin{picture}(30,20)
%
  \put( 5, 0){\line( 1, 1){20}}
  \put(25, 0){\line(-1, 1){20}}
  \put(12,20){\oval( 6,10)[rb]}
  \put(12,10){\oval( 6,10)[l]}
  \put(12, 0){\oval( 6,10)[rt]}
 \end{picture}
 \end{array}
\end{eqnarray}
in the knot theory if they are fixed
the indices upper or lower position.
A line of color-$a$ is $a$ pieces of lines that anti-symmetrized:
\begin{eqnarray}
\begin{array}{l}
\begin{picture}(20,20)
%
 \put( 8, 0){\line(0,1){20}}
 \put(12,10){\scriptsize $a$}
\end{picture}
\end{array}
 =
\begin{array}{l}
\begin{picture}(30,30)
%
 \put( 0,18){\framebox(30, 4)}
 \put( 5,10){\line(0,1){8}}
 \put(25,10){\line(0,1){8}}
 \put( 5,22){\line(0,1){8}}
 \put(25,22){\line(0,1){8}}
 \put( 9,10){$\cdots$}
 \put( 9,23){$\cdots$}
 \put( 9, 5){\oval(12,4)[tr]}
 \put(21, 5){\oval(12,4)[tl]}
 \put( 9, 9){\oval(12,4)[bl]}
 \put(21, 9){\oval(12,4)[br]}
 \put(13, 0){\scriptsize $a$}
 \end{picture}
\end{array}
 = \frac{1}{a!}\left(
\begin{array}{l}
\begin{picture}(25,20)
%
 \put( 2, 0){\line(0,1){20}}
 \put( 6, 0){\line(0,1){20}}
 \put( 9, 8){\scriptsize $\cdots$}
 \put(22, 0){\line(0,1){20}}
\end{picture}
\end{array}
 -
\begin{array}{l}
\begin{picture}(25,20)
%
 \put( 2, 0){\line( 1,5){4}}
 \put( 6, 0){\line(-1,5){4}}
 \put( 9, 8){\scriptsize $\cdots$}
 \put(22, 0){\line(0,1){20}}
\end{picture}
\end{array}
 +-\cdots
 \right),
\end{eqnarray}
where the white box means anti-symmetrizing of lines.
A trivalent vertex defined as
\begin{eqnarray}
 \begin{array}{c}
 \begin{picture}(40,25)
%
  \put(20, 3){\line(0,1){20}}
  \put( 0, 3){\line(1,0){40}}
  \put(20, 3){\circle*{3}}
  \put(22,20){\scriptsize $a$}
  \put( 5, 5){\scriptsize $b$}
  \put(30, 5){\scriptsize $c$}
 \end{picture}
 \end{array}
 =
 \begin{array}{c}
 \begin{picture}(60,41)
%
  \put(30,27){\line(0,1){14}}
  \put( 0,10){\line(1,0){13}}
  \put(17, 8){\line(1,0){26}}
  \put(47,10){\line(1,0){13}}
  \put(13, 4){\framebox( 4,12)}
  \put(43, 4){\framebox( 4,12)}
  \put(24,23){\framebox(12, 4)}
  \put(17.5,22.5){\oval(20,20)[br]}
  \put(42.5,22.5){\oval(20,20)[bl]}
  \put(23,36){\scriptsize $a$}
  \put( 5, 3){\scriptsize $b$}
  \put(50, 3){\scriptsize $c$}
  \put(26, 1){\scriptsize $i$}
  \put(40,20){\scriptsize $j$}
  \put(17,20){\scriptsize $k$}
 \end{picture}
 \end{array}\ ,\ \
\left\{
 \begin{array}{l}
  i=\frac{1}{2}(-a+b+c)\\
  j=\frac{1}{2}(a-b+c)\\
  k=\frac{1}{2}(a+b-c)
 \end{array}
\right. .
\end{eqnarray}
The generator of $su(2)$ anti-hermite 2-dimensional representation is
\begin{eqnarray}
 \tau^{i} =
 \begin{array}{l}
 \begin{picture}(17,30)
%
  \put(8,15){\circle{10}}
  \put(6,12){{\scriptsize $i$}}
  \put(8, 0){\line(0,1){10}}
  \put(8,20){\line(0,1){10}}
 \end{picture}
 \end{array} =
 \begin{array}{l}
  \begin{picture}(45,30)
%
  \put( 8,14){\circle{10}}
  \put( 6,11){{\scriptsize $i$}}
  \put(12,19){\oval(8,10)[tl]}
  \put(12, 9){\oval(8,10)[bl]}
  \put(12,14){\oval(12,20)[r]}
  \put(18,15){\line(1,0){17}}
  \put(18,15){\circle*{3}}
  \put(24,17){{\scriptsize 2}}
  \put(35,15){\circle*{3}}
  \put(35,0){\line(0,1){30}}
 \end{picture}
 \end{array}.
\end{eqnarray}
This generator acts to a line of color-$a$ as
\begin{eqnarray}
 \begin{array}{l}
 \begin{picture}(55,30)
%
  \put( 8,14){\circle{10}}
  \put( 6,11){{\scriptsize $i$}}
  \put(12,19){\oval(8,10)[tl]}
  \put(12, 9){\oval(8,10)[bl]}
  \put(12,14){\oval(12,20)[r]}
  \put(18,15){\line(1,0){17}}
  \put(18,15){\circle*{3}}
  \put(24,17){{\scriptsize 2}}
  \put(45, 0){\line(0,1){30}}
  \put(48, 9){{\scriptsize $a$}}
 \end{picture}
 \end{array}
 = a\cdot
 \begin{array}{l}
 \begin{picture}(50,30)
%
  \put( 8,14){\circle{10}}
  \put( 6,11){{\scriptsize $i$}}
  \put(12,19){\oval(8,10)[tl]}
  \put(12, 9){\oval(8,10)[bl]}
  \put(12,14){\oval(12,20)[r]}
  \put(18,15){\line(1,0){17}}
  \put(18,15){\circle*{3}}
  \put(24,17){{\scriptsize 2}}
  \put(35,15){\circle*{3}}
  \put(35, 0){\line(0,1){30}}
  \put(38, 5){{\scriptsize $a$}}
  \put(38,22){{\scriptsize $a$}}
 \end{picture}
 \end{array}.
\end{eqnarray}
Summation of product of generators
\begin{eqnarray}
 \sum_{i=1}^{3}
 \begin{array}{l}
 \begin{picture}(40,50)
%
  \put( 8,12){\circle{10}}
  \put( 6, 9){{\scriptsize $i$}}
  \put(12,17){\oval(8,10)[tl]}
  \put(12, 7){\oval(8,10)[bl]}
  \put(12,12){\oval(12,20)[r]}
  \put(18,13){\line(1,0){17}}
  \put(18,13){\circle*{3}}
  \put(24,15){{\scriptsize 2}}
%
  \put( 8,37){\circle{10}}
  \put( 6,34){{\scriptsize $i$}}
  \put(12,42){\oval(8,10)[tl]}
  \put(12,32){\oval(8,10)[bl]}
  \put(12,37){\oval(12,20)[r]}
  \put(18,38){\line(1,0){17}}
  \put(18,38){\circle*{3}}
  \put(24,40){{\scriptsize 2}}
 \end{picture}
 \end{array}
 =\frac{1}{2}\cdot
 \begin{array}{c}
 \begin{picture}(20,25)
%
  \put(12,10){\oval(20,16)[l]}
  \put( 2,18){\scriptsize 2}
 \end{picture}
 \end{array}
 \ ,\hspace{2em}
 \sum_{i,j,k}\epsilon^{ijk}
 \begin{array}{l}
 \begin{picture}(40,75)
%
  \put( 8,12){\circle{10}}
  \put( 6, 9){{\scriptsize $k$}}
  \put(12,17){\oval(8,10)[tl]}
  \put(12, 7){\oval(8,10)[bl]}
  \put(12,12){\oval(12,20)[r]}
  \put(18,13){\line(1,0){17}}
  \put(18,13){\circle*{3}}
  \put(24,15){{\scriptsize 2}}
%
  \put( 8,37){\circle{10}}
  \put( 6,35){{\scriptsize $j$}}
  \put(12,42){\oval(8,10)[tl]}
  \put(12,32){\oval(8,10)[bl]}
  \put(12,37){\oval(12,20)[r]}
  \put(18,38){\line(1,0){17}}
  \put(18,38){\circle*{3}}
  \put(24,40){{\scriptsize 2}}
%
  \put( 8,62){\circle{10}}
  \put( 6,59){{\scriptsize $i$}}
  \put(12,67){\oval(8,10)[tl]}
  \put(12,57){\oval(8,10)[bl]}
  \put(12,62){\oval(12,20)[r]}
  \put(18,63){\line(1,0){17}}
  \put(18,63){\circle*{3}}
  \put(24,65){{\scriptsize 2}}
 \end{picture}
 \end{array}
 =-\frac{1}{2}\cdot
 \begin{array}{c}
 \begin{picture}(20,45)
%
  \put(20,20){\oval(40,30)[l]}
  \put(0,20){\line(1,0){20}}
  \put(0,20){\circle*{3}}
  \put(13,38){{\scriptsize 2}}
  \put(13,23){{\scriptsize 2}}
  \put(13,8){{\scriptsize 2}}
 \end{picture}
 \end{array}.
 \label{su_two_generator}
\end{eqnarray}
The symmetrizer
\begin{eqnarray}
 \Delta_{a}&=&
 \begin{array}{c}
 \begin{picture}(20,16)
%
  \put(7,8){\oval(10,16)}
  \put(14, 4){{\scriptsize $a$}}
 \end{picture}
 \end{array}
 =(-1)^{a}(a+1).
 \label{thesymmetrizer}
\end{eqnarray}
The $\theta$-net
\begin{eqnarray}
 \theta(a,b,c)&=&
 \begin{array}{c}
 \begin{picture}(30,30)
%
  \put(15,12){\oval(30,20)}
  \put( 0,12){\line(1,0){30}}
  \put( 0,12){\circle*{3}}
  \put(30,12){\circle*{3}}
  \put(14, 3){{\scriptsize $c$}}
  \put(14,13){{\scriptsize $b$}}
  \put(14,23){{\scriptsize $a$}}
 \end{picture}
 \end{array}
 = \frac{(-1)^{m+n+p}(m+n+p+1)!m!n!p!}{a!b!c!},
 \label{thethetanet}
\end{eqnarray}
where\
$
 m=\frac{1}{2}(-a+b+c),n=\frac{1}{2}(a-b+c)
$\
and $p=\frac{1}{2}(a+b-c)$.
The tetrahedral net
\begin{eqnarray}
 \mbox{Tet}\left[
 \begin{array}{ccc}
  a&b&e\\
  c&d&f
 \end{array}
 \right]&=&
 \begin{array}{c}
 \begin{picture}(55,40)
%
  \put( 0,20){\line( 1, 0){40}}
  \put( 0,20){\line( 1, 1){20}}
  \put(40,20){\line(-1, 1){20}}
  \put( 0,20){\line( 1,-1){20}}
  \put(40,20){\line(-1,-1){20}}
  \put(20, 0){\line( 1, 0){20}}
  \put(20,40){\line( 1, 0){20}}
  \put( 0,20){\circle*{3}}
  \put(40,20){\circle*{3}}
  \put(20, 0){\circle*{3}}
  \put(20,40){\circle*{3}}
  \put(40,20){\oval(30,40)[r]}
  \put( 4, 7){{\scriptsize $a$}}
  \put( 4,29){{\scriptsize $b$}}
  \put(32,29){{\scriptsize $c$}}
  \put(32, 7){{\scriptsize $d$}}
  \put(48,15){{\scriptsize $e$}}
  \put(18,12){{\scriptsize $f$}}
 \end{picture}
 \end{array}
 =\frac{\cal I}{\cal E}\sum_{m\leq S\leq M}
  \frac{(-1)^{S}(S+1)!}{\prod_{i}(S-a_{i})!\cdot\prod_{i}(b_{i}-S)!},\\
 a_{1}&=&(a+d+e)/2,\ \
 b_{1}=(b+d+e+f)/2,\nonumber\\
 a_{2}&=&(b+c+e)/2\; ,\ \
 b_{2}=(a+c+e+f)/2,\nonumber\\
 a_{3}&=&(a+b+f)/2,\ \
 b_{3}=(a+b+c+d)/2\; ,\nonumber\\
 a_{4}&=&(c+d+f)/2,\nonumber\\
 m&=&\mbox{max}\{a_{i}\}\hspace{2em},\ \
 M=\mbox{min}\{b_{i}\}\nonumber\\
 {\cal E}&=&a!b!c!d!e!f!\hspace{1.3em} ,\ \
 {\cal I}=\mbox{$\prod_{ij}$}(b_{j}-a_{i})!.\nonumber
\end{eqnarray}
The 9-$j$ symbol
\begin{eqnarray}
 \left\{
 \begin{array}{ccc}
  a&b&c\\
  d&e&f\\
  g&h&i
 \end{array}
 \right\}&=&
 \begin{array}{c}
 \begin{picture}(85,55)
%
  \put(40,20){\line(1,0){20}}
  \put(40,50){\line(1,0){20}}
  \put( 5,18){\line(0,1){4}}
  \put(45,20){\line(0,1){30}}
  \put(35, 3){\line(0,1){15}}
  \put(45,35){\oval(60,30)}
  \put(35,18){\oval(60,30)[b]}
  \put(15,22){\oval(20,26)[tl]}
  \put(45,22){\oval(20,26)[tl]}
  \put(75,22){\oval(20,26)[tl]}
  \put(15,35){\circle*{3}}
  \put(45,20){\circle*{3}}
  \put(45,35){\circle*{3}}
  \put(45,50){\circle*{3}}
  \put(75,35){\circle*{3}}
  \put(35, 3){\circle*{3}}
  \put(10,42){\scriptsize $a$}
  \put(40,42){\scriptsize $b$}
  \put(66,42){\scriptsize $c$}
  \put(22,25){\scriptsize $d$}
  \put(47,25){\scriptsize $e$}
  \put(76,25){\scriptsize $f$}
  \put(12, 9){\scriptsize $g$}
  \put(37, 7){\scriptsize $h$}
  \put(67, 7){\scriptsize $i$}
 \end{picture}
 \end{array}.
\end{eqnarray}
The exchanging of lines in a trivalent vertex
\begin{eqnarray}
 \begin{array}{c}
 \begin{picture}(30,30)
%
  \put(15, 0){\line( 0,1){10}}
  \put( 9,16){\line( 1,1){14}}
  \bezier{20}(21,16)(19,18)(17,20)
  \bezier{25}(13,24)(10,27)( 7,30)
  \put(15,16){\oval(12,12)[b]}
  \put(15,10){\circle*{3}}
  \put( 2,25){\scriptsize $a$}
  \put(25,25){\scriptsize $b$}
  \put(18, 1){\scriptsize $c$}
 \end{picture}
 \end{array}
 &=&\lambda^{ab}_{c}
 \begin{array}{c}
 \begin{picture}(30,30)
%
  \put(15, 0){\line( 0,1){10}}
  \put(15,10){\line( 1,1){10}}
  \put(15,10){\line(-1,1){10}}
  \put(15,10){\circle*{3}}
  \put( 2,23){{\scriptsize $a$}}
  \put(24,23){{\scriptsize $b$}}
  \put(18, 3){{\scriptsize $c$}}
 \end{picture}
 \end{array}\ ,\ \
 \lambda^{ab}_{c}=(-1)^{\frac{1}{2}(a(a+3)+b(b+3)-c(c+3))}.
\end{eqnarray}
The recoupling theorem
\begin{eqnarray}
 \begin{array}{c}
 \begin{picture}(30,30)
%
  \put(10,15){\line( 1, 0){10}}
  \put(10,15){\line(-1, 1){10}}
  \put(10,15){\line(-1,-1){10}}
  \put(20,15){\line( 1, 1){10}}
  \put(20,15){\line( 1,-1){10}}
  \put(10,15){\circle*{3}}
  \put(20,15){\circle*{3}}
  \put( 2, 2){{\scriptsize $a$}}
  \put( 4,24){{\scriptsize $b$}}
  \put(22,24){{\scriptsize $c$}}
  \put(22, 2){{\scriptsize $d$}}
  \put(12, 7){{\scriptsize $f$}}
 \end{picture}
 \end{array}
 &=&\sum_{e}\left\{
 \begin{array}{ccc}
  a&b&e\\
  c&d&f
 \end{array}
 \right\}
 \begin{array}{c}
 \begin{picture}(30,30)
%
  \put(15,10){\line( 0, 1){10}}
  \put(15,10){\line( 1,-1){10}}
  \put(15,10){\line(-1,-1){10}}
  \put(15,20){\line( 1, 1){10}}
  \put(15,20){\line(-1, 1){10}}
  \put(15,10){\circle*{3}}
  \put(15,20){\circle*{3}}
  \put( 0, 2){{\scriptsize $a$}}
  \put( 0,24){{\scriptsize $b$}}
  \put(25,24){{\scriptsize $c$}}
  \put(25, 2){{\scriptsize $d$}}
  \put( 8,13){{\scriptsize $e$}}
 \end{picture}
 \end{array}.
\end{eqnarray}
The relation between a 6-$j$ symbol and a tetrahedral net
\begin{eqnarray}
 \left\{
 \begin{array}{ccc}
  a&b&e\\
  c&d&f
 \end{array}
 \right\}&=&\frac{\Delta_{e}}{\theta(a,d,e)\theta(b,c,e)}
 \mbox{Tet}\left[
 \begin{array}{ccc}
  a&b&e\\
  c&d&f
 \end{array}
 \right].
\end{eqnarray}
The reduction formulas
\begin{eqnarray}
 \begin{array}{c}
 \begin{picture}(40,40)
%
  \put(20,20){\circle{20}}
  \put(20, 0){\line( 0, 1){10}}
  \put(20,30){\line( 0, 1){10}}
  \put(20,10){\circle*{3}}
  \put(20,30){\circle*{3}}
  \put(13, 2){{\scriptsize $a$}}
  \put( 3,17){{\scriptsize $b$}}
  \put(32,17){{\scriptsize $c$}}
  \put(13,34){{\scriptsize $d$}}
 \end{picture}
 \end{array}
 &=&\delta_{ad}\frac{\theta(a,b,c)}{\Delta_{a}}
 \begin{array}{c}
 \begin{picture}(10,30)
%
  \put( 3, 0){\line( 0, 1){30}}
  \put( 5,12){{\scriptsize $a$}}
 \end{picture}
 \end{array},\\
 \begin{array}{c}
 \begin{picture}(40,40)
%
  \put(20, 5){\line(0,1){30}}
  \put(20, 5){\line(1,0){10}}
  \put(20,20){\line(1,0){10}}
  \put(20,35){\line(1,0){10}}
  \put(20,20){\oval(28,30)[l]}
  \put(20, 5){\circle*{3}}
  \put(20,20){\circle*{3}}
  \put(20,35){\circle*{3}}
  \put(33,32){\scriptsize $a$}
  \put(33, 4){\scriptsize $b$}
  \put(13,12){\scriptsize $c$}
  \put(13,25){\scriptsize $d$}
  \put( 0,18){\scriptsize $e$}
  \put(33,17){\scriptsize $f$}
 \end{picture}
 \end{array}
 &=&\frac{\mbox{Tet}\left[
 \begin{array}{ccc}
  a & b & e\\
  c & d & f
 \end{array}
 \right]}{\theta(a,f,b)}
 \begin{array}{c}
 \begin{picture}(25,30)
%
  \put( 5,15){\line(1,0){10}}
  \put(15,15){\oval(20,20)[l]}
  \put( 5,15){\circle*{3}}
  \put(17,24){\scriptsize $a$}
  \put(17,13){\scriptsize $f$}
  \put(17, 2){\scriptsize $b$}
 \end{picture}
 \end{array}.
\end{eqnarray}

\subsection{Matrix elements of volume operator}

The volume operator for the vertex $v$ is
\begin{eqnarray}
 \hat{V}=\frac{(\hbar\kappa)^{3/2}}{4}
  \sqrt{\sum_{I<J<K}\left|i\hat{W}_{[IJK]}\right|}
 \label{volume}
\end{eqnarray}
where\
$
 |\hat{X}|=\sqrt{\hat{X}^{\dagger}\hat{X}}
$.
The meaning of the square root is that
its diagonalized matrix equivalent to
square root of $\hat{X}^{\dagger}\hat{X}$.
The summation $I,J,K$ is the label of edges $e_{I}$\
connected with the vertex $v$.
The operator $\hat{W}_{[IJK]}$ adding the three $su(2)$ generators:
\begin{eqnarray}
 \hat{W}_{[IJK]}=
 \begin{array}{c}
 \begin{picture}(85,45)
%
  \put(20, 0){\line(1,0){30}}
  \put(35, 0){\line(0,1){20}}
  \put(10,10){\line(0,1){10}}
  \put(60,10){\line(0,1){10}}
  \put(20,10){\oval(20,20)[bl]}
  \put(50,10){\oval(20,20)[br]}
  \put(20,20){\oval(20,20)[tl]}
  \put(45,20){\oval(20,20)[tl]}
  \put(70,20){\oval(20,20)[tl]}
  \put(35, 0){\circle*{3}}
  \put(17,33){\scriptsize $(e_{I})$}
  \put(42,33){\scriptsize $(e_{J})$}
  \put(67,33){\scriptsize $(e_{K})$}
  \put( 2,10){\scriptsize 2}
  \put(27,10){\scriptsize 2}
  \put(52,10){\scriptsize 2}
 \end{picture}
 \end{array}
\end{eqnarray}
(times minus two) to the edges $e_{I},e_{J},e_{K}$\
(see (\ref{su_two_generator})).
Let us compute this operation to the $n$-valent vertex.
Its matrix element is
\begin{eqnarray}
 \widetilde{W}_{[IJK]\;\vec{i}}{}^{\vec{k}}&=&
 (\tilde{\phi}_{\vec{k}},\;\hat{W}_{[IJK]}\; \tilde{\phi}_{\vec{i}})
  \nonumber\\
 &=&N_{\vec{i}}N_{\vec{k}}P_{I}P_{J}P_{K}\cdot
 \begin{array}{c}
 \begin{picture}(180,80)
%
  \put( 40,30){\line(1,0){20}}
  \put( 40,60){\line(1,0){20}}
  \put( 75,30){\line(1,0){20}}
  \put( 75,60){\line(1,0){20}}
  \put(110,30){\line(1,0){20}}
  \put(110,60){\line(1,0){20}}
  \put( 50,30){\line(0,1){30}}
  \put( 85,30){\line(0,1){30}}
  \put(120,30){\line(0,1){30}}
  \put( 25,45){\oval(20,30)[l]}
  \put(145,45){\oval(20,30)[r]}
  \put( 50,30){\circle*{3}}
  \put( 50,45){\circle*{3}}
  \put( 50,60){\circle*{3}}
  \put( 85,30){\circle*{3}}
  \put( 85,45){\circle*{3}}
  \put( 85,60){\circle*{3}}
  \put(120,30){\circle*{3}}
  \put(120,45){\circle*{3}}
  \put(120,60){\circle*{3}}
  \put(  2,42){{\scriptsize $P_{0}$}}
  \put( 53,35){{\scriptsize $P_{I}$}}
  \put( 53,50){{\scriptsize $P_{I}$}}
  \put( 88,35){{\scriptsize $P_{J}$}}
  \put( 88,50){{\scriptsize $P_{J}$}}
  \put(123,35){{\scriptsize $P_{K}$}}
  \put(123,50){{\scriptsize $P_{K}$}}
  \put(158,42){{\scriptsize $P_{n-1}$}}
  \put( 33,23){{\scriptsize $i_{I}$}}
  \put( 33,63){{\scriptsize $k_{I}$}}
  \put( 52,23){{\scriptsize $i_{I+1}$}}
  \put( 52,63){{\scriptsize $k_{I+1}$}}
  \put( 68,32){{\scriptsize $i_{J}$}}
  \put( 75,63){{\scriptsize $k_{J}$}}
  \put( 88,23){{\scriptsize $i_{J+1}$}}
  \put( 88,63){{\scriptsize $k_{J+1}$}}
  \put(101,32){{\scriptsize $i_{K}$}}
  \put(108,63){{\scriptsize $k_{K}$}}
  \put(123,23){{\scriptsize $i_{K+1}$}}
  \put(123,63){{\scriptsize $k_{K+1}$}}
  \put(51,33){\oval(18,24)[tl]}
  \put(86,33){\oval(18,24)[tl]}
  \put(121,33){\oval(18,24)[tl]}
  \put(77,26){\oval(70,24)[b]}
  \put(77,14){\line(0,1){12}}
  \put(77,14){\circle*{3}}
  \put(69, 7){{\scriptsize $2$}}
  \put(81, 7){{\scriptsize $2$}}
  \put(80,17){{\scriptsize $2$}}
 \end{picture}
 \end{array}.
 \label{before}
\end{eqnarray}
We can calculate the graphical part of (\ref{before}).
Now we separate it four part (i) $e_{0}$-$e_{I}$,
(ii) $e_{I}$-$e_{J}$, (iii) $e_{J}$-$e_{K}$ and
(iv) $e_{K}$-$e_{n-1}$.

(i)
\begin{eqnarray}
 \begin{array}{c}
 \begin{picture}(75,65)
  \put(25,35){\oval(20,30)[l]}
  \put(40,20){\line(1,0){20}}
  \put(40,50){\line(1,0){20}}
  \put(50,20){\line(0,1){30}}
  \put(50,20){\circle*{3}}
  \put(50,35){\circle*{3}}
  \put(50,50){\circle*{3}}
  \put( 2,32){{\scriptsize $P_{0}$}}
  \put(53,25){{\scriptsize $P_{I}$}}
  \put(53,40){{\scriptsize $P_{I}$}}
  \put(33,13){{\scriptsize $i_{I}$}}
  \put(33,53){{\scriptsize $k_{I}$}}
  \put(52,13){{\scriptsize $i_{I+1}$}}
  \put(52,53){{\scriptsize $k_{I+1}$}}
  \put(51,23){\oval(18,24)[tl]}
  \put(60,16){\oval(36,24)[bl]}
  \put(65, 0){{\scriptsize $2$}}
 \end{picture}
 \end{array}
 &=&\lambda^{P_{I}2}_{P_{I}}\cdot\!
 \begin{array}{c}
 \begin{picture}(92,65)
%
  \put(25,35){\oval(20,30)[l]}
  \put(50,23){\oval(18,24)[tr]}
  \put(70,16){\oval(22,24)[bl]}
  \put(40,20){\line(1,0){30}}
  \put(40,50){\line(1,0){30}}
  \put(50,20){\line(0,1){30}}
  \put(50,20){\circle*{3}}
  \put(50,35){\circle*{3}}
  \put(50,50){\circle*{3}}
  \put( 2,32){{\scriptsize $P_{0}$}}
  \put(38,25){{\scriptsize $P_{I}$}}
  \put(53,40){{\scriptsize $P_{I}$}}
  \put(33,13){{\scriptsize $i_{I}$}}
  \put(33,53){{\scriptsize $k_{I}$}}
  \put(73,18){{\scriptsize $i_{I+1}$}}
  \put(52,53){{\scriptsize $k_{I+1}$}}
  \put(73, 2){{\scriptsize $2$}}
 \end{picture}
 \end{array}
  \nonumber\\
 &=&
 -\lambda^{i_{I+1}2}_{k_{I+1}}
  (\prod_{x=2}^{r}\delta_{i_{x}}^{k_{x}})
  \frac{\theta(P_{0},P_{1},i_{2})}{\Delta_{i_{2}}}
  (\prod_{x=2}^{r-1}\frac{\theta(i_{x},P_{x},i_{x+1})}{\Delta_{i_{x+1}}})
   \nonumber\\
 &&\times
   \frac{\mbox{Tet}\left[
  \begin{array}{ccc}
   k_{I+1} & i_{I+1} & i_{I} \\
   P_{I}   & P_{I}   & 2
  \end{array}
  \right]}{\theta(k_{I+1},i_{I+1},2)}\cdot\!
 \begin{array}{c}
 \begin{picture}(55,65)
%
  \put(30,35){\oval(20,30)[l]}
  \put(20,20){\oval(18,30)[l]}
  \put(20,5){\line(1,0){10}}
  \put(20,35){\circle*{3}}
  \put(33,15){{\scriptsize $i_{I+1}$}}
  \put(33,53){{\scriptsize $k_{I+1}$}}
  \put(32, 2){{\scriptsize $2$}}
 \end{picture}
 \end{array}\\
 &=& \alpha_{\mbox{\scriptsize i}}\cdot
 \begin{array}{c}
 \begin{picture}(55,65)
%
  \put(30,35){\oval(20,30)[l]}
  \put(20,20){\oval(18,30)[l]}
  \put(20,5){\line(1,0){10}}
  \put(20,35){\circle*{3}}
  \put(33,15){{\scriptsize $i_{I+1}$}}
  \put(33,53){{\scriptsize $k_{I+1}$}}
  \put(32, 2){{\scriptsize $2$}}
 \end{picture}
 \end{array}
\end{eqnarray}

(iv)
\begin{eqnarray}
 \begin{array}{c}
 \begin{picture}(80,55)
%
  \put(45,25){\oval(20,30)[r]}
  \put(10,10){\line(1,0){20}}
  \put(10,40){\line(1,0){20}}
  \put(20,10){\line(0,1){30}}
  \put(20,10){\circle*{3}}
  \put(20,25){\circle*{3}}
  \put(20,40){\circle*{3}}
  \put(58,23){{\scriptsize $P_{n-1}$}}
  \put(23,15){{\scriptsize $P_{K}$}}
  \put(23,30){{\scriptsize $P_{K}$}}
  \put( 0, 8){{\scriptsize $i_{K}$}}
  \put( 3,23){{\scriptsize $2$}}
  \put( 3,43){{\scriptsize $k_{K}$}}
  \put(22, 3){{\scriptsize $i_{K+1}$}}
  \put(22,43){{\scriptsize $k_{K+1}$}}
  \put(10,25){\line(1,0){10}}
 \end{picture}
 \end{array}
 &=&
 (\prod_{x=t+1}^{n-2}\delta_{i_{x}}^{k_{x}})
  (\prod_{x=t+1}^{n-3}\frac{\theta(i_{x},P_{x},i_{x+1})
   }{\Delta_{i_{x}}})
  \frac{\theta(i_{n-2},P_{n-2},P_{n-1})}{\Delta_{i_{n-2}}}
  \nonumber\\
 &&\times
  \frac{\mbox{Tet}\left[
  \begin{array}{ccc}
   i_{K} & k_{K} & i_{K+1} \\
   P_{K} & P_{K} & 2
  \end{array}
  \right]}{\theta(k_{K},i_{K},2)}
 \cdot
 \begin{array}{c}
 \begin{picture}(45,50)
  \put(25,25){\oval(20,30)[r]}
  \put(15,40){\line(1,0){10}}
  \put(15,25){\line(1,0){20}}
  \put(15,10){\line(1,0){10}}
  \put(35,25){\circle*{3}}
  \put( 5, 7){{\scriptsize $i_{K}$}}
  \put( 5,22){{\scriptsize $2$}}
  \put( 3,37){{\scriptsize $k_{K}$}}
 \end{picture}
 \end{array}\\
 &=&\alpha_{\mbox{\scriptsize iv}}\cdot
 \begin{array}{c}
 \begin{picture}(45,50)
  \put(25,25){\oval(20,30)[r]}
  \put(15,40){\line(1,0){10}}
  \put(15,25){\line(1,0){20}}
  \put(15,10){\line(1,0){10}}
  \put(35,25){\circle*{3}}
  \put( 5, 7){{\scriptsize $i_{K}$}}
  \put( 5,22){{\scriptsize $2$}}
  \put( 3,37){{\scriptsize $k_{K}$}}
 \end{picture}
 \end{array}
\end{eqnarray}
Then the graph of (\ref{before}) is
\begin{eqnarray}
 \begin{array}{c}
 \begin{picture}(180,75)
%
  \put( 40,30){\line(1,0){20}}
  \put( 40,60){\line(1,0){20}}
  \put( 75,30){\line(1,0){20}}
  \put( 75,60){\line(1,0){20}}
  \put(110,30){\line(1,0){20}}
  \put(110,60){\line(1,0){20}}
  \put( 50,30){\line(0,1){30}}
  \put( 85,30){\line(0,1){30}}
  \put(120,30){\line(0,1){30}}
  \put( 25,45){\oval(20,30)[l]}
  \put(145,45){\oval(20,30)[r]}
  \put( 50,30){\circle*{3}}
  \put( 50,45){\circle*{3}}
  \put( 50,60){\circle*{3}}
  \put( 85,30){\circle*{3}}
  \put( 85,45){\circle*{3}}
  \put( 85,60){\circle*{3}}
  \put(120,30){\circle*{3}}
  \put(120,45){\circle*{3}}
  \put(120,60){\circle*{3}}
  \put(  2,42){{\scriptsize $P_{0}$}}
  \put( 53,35){{\scriptsize $P_{I}$}}
  \put( 53,50){{\scriptsize $P_{I}$}}
  \put( 88,35){{\scriptsize $P_{J}$}}
  \put( 88,50){{\scriptsize $P_{J}$}}
  \put(123,35){{\scriptsize $P_{K}$}}
  \put(123,50){{\scriptsize $P_{K}$}}
  \put(158,42){{\scriptsize $P_{n-1}$}}
  \put( 33,23){{\scriptsize $i_{I}$}}
  \put( 33,63){{\scriptsize $k_{I}$}}
  \put( 52,23){{\scriptsize $i_{I+1}$}}
  \put( 52,63){{\scriptsize $k_{I+1}$}}
  \put( 68,32){{\scriptsize $i_{J}$}}
  \put( 75,63){{\scriptsize $k_{J}$}}
  \put( 88,23){{\scriptsize $i_{J+1}$}}
  \put( 88,63){{\scriptsize $k_{J+1}$}}
  \put(101,32){{\scriptsize $i_{K}$}}
  \put(108,63){{\scriptsize $k_{K}$}}
  \put(123,23){{\scriptsize $i_{K+1}$}}
  \put(123,63){{\scriptsize $k_{K+1}$}}
  \put( 51,33){\oval(18,24)[tl]}
  \put( 86,33){\oval(18,24)[tl]}
  \put(121,33){\oval(18,24)[tl]}
  \put( 77,26){\oval(70,24)[b]}
  \put( 77,14){\line(0,1){12}}
  \put( 77,14){\circle*{3}}
  \put( 69, 7){{\scriptsize $2$}}
  \put( 81, 7){{\scriptsize $2$}}
  \put( 80,17){{\scriptsize $2$}}
 \end{picture}
 \end{array}
 &=&\alpha_{\mbox{\scriptsize i}}\alpha_{\mbox{\scriptsize iv}}\cdot\hspace{-1em}
 \begin{array}{c}
 \begin{picture}(110,75)
%
  \put(45,30){\line(1,0){20}}
  \put(45,60){\line(1,0){20}}
  \put(12,26){\line(0,1){7}}
  \put(55,30){\line(0,1){30}}
  \put(30,45){\oval(20,30)[l]}
  \put(80,45){\oval(20,30)[r]}
  \put(20,45){\circle*{3}}
  \put(55,30){\circle*{3}}
  \put(55,45){\circle*{3}}
  \put(55,60){\circle*{3}}
  \put(90,45){\circle*{3}}
  \put(58,35){{\scriptsize $P_{J}$}}
  \put(58,50){{\scriptsize $P_{J}$}}
  \put(22,23){{\scriptsize $i_{I+1}$}}
  \put(22,63){{\scriptsize $k_{I+1}$}}
  \put(38,32){{\scriptsize $i_{J}$}}
  \put(45,63){{\scriptsize $k_{J}$}}
  \put(58,23){{\scriptsize $i_{J+1}$}}
  \put(58,63){{\scriptsize $k_{J+1}$}}
  \put(92,30){{\scriptsize $i_{K}$}}
  \put(88,60){{\scriptsize $k_{K}$}}
  \put(21,33){\oval(18,24)[tl]}
  \put(56,33){\oval(18,24)[tl]}
  \put(91,33){\oval(18,24)[tl]}
  \put(47,26){\oval(70,24)[b]}
  \put(47,14){\line(0,1){12}}
  \put(47,14){\circle*{3}}
  \put(39, 7){{\scriptsize $2$}}
  \put(51, 7){{\scriptsize $2$}}
  \put(50,17){{\scriptsize $2$}}
 \end{picture}
 \end{array}.
\end{eqnarray}
Next, we calculate the remaining parts (ii) and (iii).

(ii)
\begin{eqnarray}
 \begin{array}{c}
 \begin{picture}(90,60)
%
  \put(30,30){\oval(20,30)[l]}
  \put(10,30){\line(1,0){10}}
  \put(50,15){\line(1,0){20}}
  \put(50,45){\line(1,0){20}}
  \put(60,15){\line(0,1){30}}
  \put(20,30){\circle*{3}}
  \put(60,15){\circle*{3}}
  \put(60,45){\circle*{3}}
  \put( 0,28){{\scriptsize $2$}}
  \put(12,48){{\scriptsize $k_{I+1}$}}
  \put(12, 9){{\scriptsize $i_{I+1}$}}
  \put(40,48){{\scriptsize $k_{J-1}$}}
  \put(40, 9){{\scriptsize $i_{J-1}$}}
  \put(70,48){{\scriptsize $k_{J}$}}
  \put(70, 7){{\scriptsize $i_{J}$}}
  \put(63,27){{\scriptsize $P_{s-1}$}}
 \end{picture}
 \end{array}
 &=&
  \left(\frac{\mathstrut}{\mathstrut}\right.
  \prod_{x=r+1}^{s-1}\frac{\mbox{Tet}\left[%
  \begin{array}{ccc}
   k_{x}   & k_{x+1} & 2\\
   i_{x+1} & i_{x}   & P_{x}
  \end{array}
  \right]}{\theta(k_{x+1},i_{x+1},2)}
  \left.\frac{\mathstrut}{\mathstrut}\right)
 \begin{array}{c}
 \begin{picture}(45,60)
%
  \put(30,30){\oval(20,30)[l]}
  \put(10,30){\line(1,0){10}}
  \put(20,30){\circle*{3}}
  \put( 0,28){{\scriptsize $2$}}
  \put(32,48){{\scriptsize $k_{J}$}}
  \put(32, 9){{\scriptsize $i_{J}$}}
 \end{picture}
 \end{array}\\
 &=&\alpha_{\mbox{\scriptsize ii}}\cdot
 \begin{array}{c}
 \begin{picture}(45,60)
%
  \put(30,30){\oval(20,30)[l]}
  \put(10,30){\line(1,0){10}}
  \put(20,30){\circle*{3}}
  \put( 0,28){{\scriptsize $2$}}
  \put(32,48){{\scriptsize $k_{J}$}}
  \put(32, 9){{\scriptsize $i_{J}$}}
 \end{picture}
 \end{array}
\end{eqnarray}

(iii)
\begin{eqnarray}
 \begin{array}{c}
 \begin{picture}(70,70)
%
  \put(50,40){\oval(20,30)[r]}
  \put(60,28){\oval(18,24)[tl]}
  \put(42,20){\oval(18,24)[br]}
  \put(60,40){\circle*{3}}
  \put(15,25){\line(1,0){20}}
  \put(15,55){\line(1,0){20}}
  \put(25,25){\line(0,1){30}}
  \put(25,25){\circle*{3}}
  \put(25,55){\circle*{3}}
  \put(35, 5){{\scriptsize $2$}}
  \put( 5,60){{\scriptsize $k_{J+1}$}}
  \put( 5,17){{\scriptsize $i_{J+1}$}}
  \put(27,60){{\scriptsize $k_{J+2}$}}
  \put(27,17){{\scriptsize $i_{J+2}$}}
  \put(63,48){{\scriptsize $k_{K}$}}
  \put(63,27){{\scriptsize $i_{K}$}}
  \put( 5,37){{\scriptsize $P_{s+1}$}}
 \end{picture}
 \end{array}
 &=&\lambda^{i_{K}2}_{k_{K}}\cdot\!
 \begin{array}{c}
 \begin{picture}(80,70)
%
  \put(50,40){\oval(20,30)[r]}
  \put(60,24){\oval(18,32)[r]}
  \put(42, 8){\line(1,0){18}}
  \put(60,40){\circle*{3}}
  \put(15,25){\line(1,0){20}}
  \put(15,55){\line(1,0){20}}
  \put(25,25){\line(0,1){30}}
  \put(25,25){\circle*{3}}
  \put(25,55){\circle*{3}}
  \put(35, 5){{\scriptsize $2$}}
  \put( 5,60){{\scriptsize $k_{J+1}$}}
  \put( 5,17){{\scriptsize $i_{J+1}$}}
  \put(27,60){{\scriptsize $k_{J+2}$}}
  \put(27,17){{\scriptsize $i_{J+2}$}}
  \put(58,53){{\scriptsize $k_{K}$}}
  \put(48,31){{\scriptsize $i_{K}$}}
  \put( 5,37){{\scriptsize $P_{s+1}$}}
 \end{picture}
 \end{array}
 \nonumber\\
 &=&\frac{\lambda^{i_{K}2}_{k_{K}}}{\lambda^{i_{J+1}2}_{k_{J+1}}}
  \left(\frac{\mathstrut}{\mathstrut}\right.
   \prod_{x=s+1}^{t-1}\frac{\mbox{Tet}\left[
   \begin{array}{ccc}
     k_{x}   & k_{x+1} & 2\\
     i_{x+1} & i_{x}   & P_{x}
   \end{array}
   \right]}{\theta(k_{x},i_{x},2)}
  \left.\frac{\mathstrut}{\mathstrut}\right)
 \begin{array}{c}
 \begin{picture}(35,70)
%
  \put(20,40){\oval(20,30)[r]}
  \put(30,28){\oval(18,24)[tl]}
  \put(12,20){\oval(18,24)[br]}
  \put(30,40){\circle*{3}}
  \put(12,25){\line(1,0){8}}
  \put(12,55){\line(1,0){8}}
  \put( 5, 5){{\scriptsize $2$}}
  \put( 5,58){{\scriptsize $k_{J+1}$}}
  \put( 3,19){{\scriptsize $i_{J+1}$}}
 \end{picture}
 \end{array}\\
 &=&\alpha_{\mbox{\scriptsize iii}}\cdot
 \begin{array}{c}
 \begin{picture}(35,70)
%
  \put(20,40){\oval(20,30)[r]}
  \put(30,28){\oval(18,24)[tl]}
  \put(12,20){\oval(18,24)[br]}
  \put(30,40){\circle*{3}}
  \put(12,25){\line(1,0){8}}
  \put(12,55){\line(1,0){8}}
  \put( 5, 5){{\scriptsize $2$}}
  \put( 5,58){{\scriptsize $k_{J+1}$}}
  \put( 3,19){{\scriptsize $i_{J+1}$}}
 \end{picture}
 \end{array}
\end{eqnarray}
Then we obtain
\begin{eqnarray}
 \widetilde{W}_{[IJK]\vec{i}}{}^{\vec{k}}&=&
  N_{\vec{i}}N_{\vec{k}}P_{I}P_{J}P_{K}\cdot
 \begin{array}{c}
 \begin{picture}(180,75)
%
  \put( 40,30){\line(1,0){20}}
  \put( 40,60){\line(1,0){20}}
  \put( 75,30){\line(1,0){20}}
  \put( 75,60){\line(1,0){20}}
  \put(110,30){\line(1,0){20}}
  \put(110,60){\line(1,0){20}}
  \put( 50,30){\line(0,1){30}}
  \put( 85,30){\line(0,1){30}}
  \put(120,30){\line(0,1){30}}
  \put( 25,45){\oval(20,30)[l]}
  \put(145,45){\oval(20,30)[r]}
  \put( 50,30){\circle*{3}}
  \put( 50,45){\circle*{3}}
  \put( 50,60){\circle*{3}}
  \put( 85,30){\circle*{3}}
  \put( 85,45){\circle*{3}}
  \put( 85,60){\circle*{3}}
  \put(120,30){\circle*{3}}
  \put(120,45){\circle*{3}}
  \put(120,60){\circle*{3}}
  \put(  2,42){{\scriptsize $P_{0}$}}
  \put( 53,35){{\scriptsize $P_{I}$}}
  \put( 53,50){{\scriptsize $P_{I}$}}
  \put( 88,35){{\scriptsize $P_{J}$}}
  \put( 88,50){{\scriptsize $P_{J}$}}
  \put(123,35){{\scriptsize $P_{K}$}}
  \put(123,50){{\scriptsize $P_{K}$}}
  \put(158,42){{\scriptsize $P_{n-1}$}}
  \put( 33,23){{\scriptsize $i_{I}$}}
  \put( 33,63){{\scriptsize $k_{I}$}}
  \put( 52,23){{\scriptsize $i_{I+1}$}}
  \put( 52,63){{\scriptsize $k_{I+1}$}}
  \put( 68,32){{\scriptsize $i_{J}$}}
  \put( 75,63){{\scriptsize $k_{J}$}}
  \put( 88,23){{\scriptsize $i_{J+1}$}}
  \put( 88,63){{\scriptsize $k_{J+1}$}}
  \put(101,32){{\scriptsize $i_{K}$}}
  \put(108,63){{\scriptsize $k_{K}$}}
  \put(123,23){{\scriptsize $i_{K+1}$}}
  \put(123,63){{\scriptsize $k_{K+1}$}}
  \put( 51,33){\oval(18,24)[tl]}
  \put( 86,33){\oval(18,24)[tl]}
  \put(121,33){\oval(18,24)[tl]}
  \put( 77,26){\oval(70,24)[b]}
  \put( 77,14){\line(0,1){12}}
  \put( 77,14){\circle*{3}}
  \put( 69, 7){{\scriptsize $2$}}
  \put( 81, 7){{\scriptsize $2$}}
  \put( 80,17){{\scriptsize $2$}}
 \end{picture}
 \end{array}\nonumber\\
 &=&
 N_{\vec{i}}N_{\vec{k}}P_{I}P_{J}P_{K}\cdot
 \alpha_{\mbox{\scriptsize i}}\alpha_{\mbox{\scriptsize ii}}
 \alpha_{\mbox{\scriptsize iii}}\alpha_{\mbox{\scriptsize iv}}\cdot
 \left\{
  \begin{array}{ccc}
   k_{J} & P_{J} & k_{J+1}\\
   i_{J} & P_{J} & i_{J+1}\\
   2     & 2     & 2
  \end{array}
 \right\}.
\end{eqnarray}

Since $i\widetilde{W}_{[IJK]}$ is a pure imaginally anti-symmetric matrix,
it can be diagonalized.
Using unitary matrix $U_{[IJK]}$ which
diagonalize it,
\begin{equation}
 \sum_{I<J<K}|i\hat{W}_{[IJK]}|\; \tilde{\phi}_{\vec{i}}=
  \sum_{I<J<K}\sum_{\vec{j},\vec{k}}
  U^{\dagger}_{[IJK]\;\vec{i}}{}^{\vec{j}}
  |i\widetilde{W}_{[IJK]\;\vec{j}}^{\mbox{\scriptsize diag}}|
  U_{[IJK]\;\vec{j}}{}^{\vec{k}}
  \;\tilde{\phi}_{\vec{k}}
\end{equation}
The sum of hermite matrices is also hermite,
(\ref{volume}) can be diagonalized.
Thus, the matrix element of $\hat{V}^{n}$ is
\begin{eqnarray}
 \hat{V}^{n}\tilde{\phi}_{\vec{i}}&=&
 \sum_{\vec{j},\vec{k}}R^{\dagger}_{\vec{i}}{}^{\vec{j}}
  ((V^{2})^{\mbox{\scriptsize diag}}_{\vec{j}})^{n/2}
  R_{\vec{j}}{}^{\vec{k}}\tilde{\phi}_{\vec{k}}\nonumber\\
 &=&
 \frac{(\hbar\kappa)^{3n/2}}{4^{n}}
  \sum_{\vec{j},\vec{k},\vec{l},\vec{m},\vec{n}}
  R^{\dagger}_{\vec{i}}{}^{\vec{j}}[
   R_{\vec{j}}{}^{\vec{k}}\sum_{I<J<K}U^{\dagger}_{[IJK]\vec{k}}{}^{\vec{l}}
   |i\widetilde{W}^{\mbox{\scriptsize diag}}_{[IJK]\vec{l}}|
   U_{[IJK]\vec{l}}{}^{\vec{m}}R^{\dagger}_{\vec{m}}{}^{\vec{j}}
  ]^{n/2}\ R_{\vec{j}}{}^{\vec{n}}\tilde{\phi}_{\vec{n}}.
 \label{volume_element}
\end{eqnarray}



\end{document}